\documentclass[a4paper,fleqn,usenatbib]{mnras}

\usepackage[T1]{fontenc}
\usepackage{ae,aecompl}

\usepackage{graphicx}	% Including figure files
\usepackage{amsmath}	% Advanced maths commands
\usepackage{amssymb}	% Extra maths symbols
\usepackage{caption}
\usepackage{subcaption}
\usepackage{subcaption}
\title[KAT-7 observations of NGC 6822]{H\sc{i} Kinematics, Mass Distribution, and Star Formation Threshold in NGC 6822, using the SKA pathfinder KAT-7}

\author[Namumba et al.]{
B.\ Namumba,$^{1}$\thanks{E-mail: brenda@ast.uct.ac.za}
C.\ Carignan, $^{1,2}$
S.\ Passmoor, $^{3}$
W.\ J.\ G.\ de Blok $^{4,1,5}$
\\
$^1$Department of Astronomy, University of Cape Town, Private Bag X3, Rondebosch 7701, South Africa\\
$^2$Observatoire d$^{\prime}$Astrophysique de l$^{\prime}$Universit$\acute{e}$ de Ouagadougou, BP 7021, Ouagadougou 03, Bukina Faso \\
$^3$SKA South Africa, The Park, Park Road, Pinelands, 7705, South Africa\\
$^{4}$ASTRON, the Netherlands Institute for Radio Astronomy, Postbus 2, NL-7990 AA Dwingeloo, The Netherlands\\
$^{5}$Kapteyn Astronomical Institute, University of Groningen, P.O Box 800, 9700 AV Groningen, The Netherlands
}

\date{Accepted 2017 August 29. Received 2017 August 25; in original form 2017 July 07}

\pubyear{2017}

\begin{document}
\label{firstpage}
\pagerange{\pageref{firstpage}--\pageref{lastpage}}
\maketitle

\begin{abstract}
We present high sensitivity H\textsc{i} observations of NGC 6822, obtained with the Karoo Array Telescope (KAT-7). We study the kinematics, the mass distribution, and the star formation thresholds. The KAT-7 short baselines and low system temperature make it sensitive to large-scale, low surface brightness emission. The observations detected $\sim$ 23$\%$ more flux than previous ATCA observations. We fit a tilted ring model to the H\textsc{i} velocity field to derive the rotation curve (RC). The KAT-7 observations allow the measurement of the rotation curve of NGC 6822 out to 5.8 kpc, $\sim$ 1 kpc further than existing measurements.

NGC 6822 is seen to be dark matter dominated at all radii. The observationally motivated pseudo-isothermal dark matter (DM) halo model reproduces well the observed RC while the Navarro Frank-White DM model gives a poor fit to the data. We find the best fit mass to light ratio (M/L) of 0.12 $\pm$ 0.01 which is consistent with the literature. The Modified Newtonian Dynamics (MOND) gives a poor fit to our data.

We derive the star formation threshold in NGC 6822 using the H\textsc{i} and H$\alpha$ data. The critical gas densities were calculated for gravitational instabilities using the Toomre-Q criterion and the cloud-growth criterion. We found that in regions of star formation, the cloud-growth criterion explains star formation better than the Toomre-Q criterion. This shows that the local shear rate could be a key player in cloud formation for irregular galaxies such as NGC 6822.
\end{abstract}


%%%%%%%%%%%%%%%%%%%%%%%%%%%%%%%%%%%%%%%%%%%%%%%%%%

%%%%%%%%%%%%%%%%% BODY OF PAPER %%%%%%%%%%%%%%%%%%

\section{Introduction}
%Dwarf irregular (dIrr) galaxies constitute a considerable fraction of all galaxies and are prevalent among the effectively star-forming systems. Their simplicity, free from spiral density waves, makes them excellent laboratories in which we can study the stellar and gas content. The most extensively studied of these dIrr galaxies are in the Local Group, as their proximity allows us detailed information of their H\textsc{i} envelopes. DIrr galaxies often show H\textsc{i} envelopes that extend significantly outwards beyond the stellar disk (e.g.,\ \citealp{1981A&A...102..134H,1998AJ....116.2363W,2007MNRAS.375..199G,2011AJ....141..204K,2014A&A...561A..28S}). These extended structures provide crucial information on the formation and evolution of the galaxies, and allow the gas dynamics of these systems to be traced at large distances from their centers (e.g., \ \citealp{1989ApJ...347..760C,1998ApJ...506..125C}), therefore probing their dark matter halos over a wide radial range.
Dwarf irregular (dIrr) galaxies constitute a considerable fraction of all galaxies and are prevalent among the effectively star-forming systems. Their simplicity, free from spiral density waves, makes them excellent laboratories in which we can study the stellar gas content. The most extensively studied of these dIrr galaxies are in the Local Group, as their proximity allows us detailed information of their H\textsc{i} envelopes. DIrr galaxies often show H\textsc{i} envelopes that extend significantly outwards beyond the stellar disk (e.g.,\ \citealp{1981A&A...102..134H,1998AJ....116.2363W,2007MNRAS.375..199G,2011AJ....141..204K,2014A&A...561A..28S}). These extended structures provide crucial information on the formation and evolution of the galaxies, and allow the gas dynamics of these systems to be traced at large distances from the centers (e.g., \ \citealp{1989ApJ...347..760C,1998ApJ...506..125C}), therefore probing their dark matter halos over a wide radial range. 

The evolution of galaxies is strongly influenced by how quickly gas is consumed by stars. Despite being gas-rich, the dIrr galaxies show very little star formation as compared to spiral galaxies. This has been one of the early puzzles about these galaxies. The reason for this is not well-understood but it could indicate inefficient star formation or that the gas density is below the threshold value. The existence of surface density thresholds for star formation is, explained in terms of the Toomre $Q$ criterion for gravitational instability which describes the gravitational instabilities in a gaseous, rotating disk \citep{1972ApJ...176L...9Q}. \citet{1989ApJ...344..685K} and \citet{2001ApJ...555..301M} explained the gravitational instabilities using the azimuthally averaged CO/H\textsc{i} and H$\alpha$ profiles, thereby providing a global picture of the threshold. \citet{2008AJ....136.2846B} used a sample of spiral and dwarf galaxies to investigate how star formation laws differ between molecular gas (H$_{2}$) centers of spiral galaxies, their H\textsc{i} dominated outskirts and the H\textsc{i} rich late-type dwarf galaxies. They found a good correlation between star formation rate density and the molecular hydrogen surface density. These results suggest that molecular gas forms stars at a constant efficiency in spirals. They found that the relationship between H\textsc{i} and star formation rate in spiral galaxies varies dramatically as a function of radius while the star formation efficiency in dwarf galaxies showed a similar trend to that found in the outer parts of spiral galaxies. They concluded that rotational shear, which is typically absent in dwarf galaxies cannot drive star formation efficiencies. In certain galaxies, star forming regions have been found in regions with gas densities below the threshold value. H$\alpha$ imaging of three nearby galaxies \citep{1998ApJ...506L..19F} revealed the presence of H\textsc{ii} regions out to, and beyond the optical radii, R$_{25}$. More recently, studies showing the bright stellar complexes in the extreme outer disk of M83, extending to about 4 times the radius at which the majority of H\textsc{ii} regions are detected suggests that H$\alpha$ may not be a complete tracer of ongoing star formation in disk galaxies \citep{2005ApJ...619L..79T}. 

Several studies of irregular galaxies tracing the extent and dynamics of the H\textsc{i} gas are mostly available from observations performed with arrays having higher resolution and lack short baselines. (e.g.,\ \citealp{1986A&A...165...45S,2000AJ....120.3027C,2000ApJ...537L..95D,2011AJ....141..204K,2012AJ....144..134H}). This means that we could be missing out on the low surface brightness extended H\textsc{i} in these galaxies, therefore underestimating their H\textsc{i} extent. To address this problem, we would ideally want to use a single dish telescope as this does not resolve out any emission, however, these dishes do not have the required spatial resolution. Our best alternative is to use an interferometer with short baselines optimized for detecting low column density regions. 

NGC 6822 is one of the most intriguing dwarf galaxies in the Local Group located at a distance of 480 $\pm$ 20 kpc \citep{2012MNRAS.421.2998F}, making it one of the closest dIrr in the Local Group after the Magellanic Cloud (SMC $\&$ LMC). It does not appear to be associated with either the Milky Way or M31, and it has no other neighboring companion. NGC 6822 is faint, with an absolute magnitude of M$_{B}$ = -15.2 \citep{1998ARA&A..36..435M}. The basic parameters of NGC 6822 are given in Table 1. The H\textsc{i} disc of this galaxy was first studied by \citet{1961BAN....15..307V}, using observations performed with the single dish Dwingeloo telescope. Their results show a disk-like system, with large amounts of H\textsc{i} (1.5 $\times 10^{8} M_{\odot}$) rotating in an orderly pattern about the center field. The first interferometry study was done by \citet{1977A&A....61..523G} who used observations from the Owens Valley Radio Observatory interferometer to study the neutral hydrogen associated with the optical core of the galaxy. Their results already showed highly structured ISM. \citet{1998AAS...193.7011B} used the VLA to study the H\textsc{i} distribution and kinematics in NGC 6822. Their H\textsc{i} images revealed a complex and clumpy interstellar medium with shell structures found over a large range of size scales. \citet{2000ApJ...537L..95D} used the low and high resolution observations from Parkes single dish and the Australian Telescope Compact Array (ATCA) to derive the H\textsc{i} kinematics and dynamics in NGC 6822. Their results show that NGC 6822 is not a quiescent non-interacting dwarf galaxy, but that it is undergoing an interaction with a possible companion. The RC obtained from the high resolution data showed that NGC 6822 is dark matter dominated \citep{2003MNRAS.340...12W}.

The proximity, isolation and lack of spiral arm structures of NGC 6822 makes it an excellent candidate to investigate star formation processes other than spiral density waves. \citet{2006AJ....131..363D} used the high resolution, wide field, deep HI, H$\alpha$ and optical data to derive the star formation thresholds in NGC 6822. Their studies revealed that the interpretation of the star formation threshold depends on the velocity dispersion used. They showed that when the velocity dispersion of 6 kms$^{-1}$ is used, the value which is consistent with previous works and the value derived from the moment analysis, the disk is stable against star formation everywhere. In this case, the star formation is attributed to the local density enhancements which would be averaged out in the radial profile, giving the impression of a stable disk. Their results show that using the velocity dispersion of the cool component ($\sim$4 kms$^{-1}$) fits the distribution of the current star formation in NGC 6822. In this case, they suggested that star formation is self regulatory. 

We present H\textsc{i} observations of NGC 6822 taken with the Karoo Array Telescope (KAT-7). With the H\textsc{i} diameter of $>$ 1 degree \citep{2000ApJ...537L..95D}, NGC 6822 is a perfect target to be observed with the KAT-7 and falls in its ``niche'': detecting low column density large scale structures not visible to other synthesis arrays such as the VLA and ATCA. The purpose of these observations is to study the distribution and kinematics of the neutral gas and the dynamics of the galaxy. Furthermore, we aim to understand the nature of dark matter associated with this galaxy at much larger radii and derive the star formation thresholds. The remainder of this paper is as follows. KAT-7 H\textsc{i} observations and data reduction are summarized in Section 2, while the H\textsc{i} results appear in Section 3. Section 4 details the derivation of the rotation curve that is used for the mass model analysis of Section 5. In Section 6, we explore instability models for the onset of star formation in NGC 6822 and in Section 7 we summarize our work.

\begin{table}
\scriptsize

\caption{\small Basic properties of NGC 6822.}
\begin{minipage}{\textwidth}
\begin{tabular}{l@{\hspace{0.30cm}}c@{\hspace{0.20cm}}c@{\hspace{0.10cm}}}   
\hline

Parameter & &  Ref \\
                         %&(sec)       &(Mpc) & (mag) & ($L_{\odot}$) &($\rm M_{\odot}yr^{-1}$)  \\
                %~~~~~~(1)    &   (2)        &  (3)   \\         
   
\hline \hline  
%\multicolumn{6}{@{} p{8.5 cm} @{}}{\footnotesize{\hspace{3cm} VLT/NACO DATA}}\\\\
Morphology Type  & IB(s)m&(a) \\
Right ascension (J2000) &19:44:57.9& (a) \\
Declination (J2000)&-14:48:11.0& (a) \\
Distance (kpc) & 480$\pm$ 20&(g)\\
Isophotal major diameter, D$_{25}$ (kpc) &2.7&(c) \\
Central surface brightness, B(0)$_{c}$ (mag/arcsec$^{2}$) & 19.8& (d)\\
Scale length, $\alpha^{-1}$(arcmin) &4.7& (d) \\
Metallicity(Z$_{\bigodot}$) &0.3&(e) \\
Total optical luminosity(L$_{V}$)(L$_{\odot}$)&$\sim $ 9.4$\times 10^{7}$& (b) \\
V$_{\text{heliocentric}}$ (kms$^{-1}$) & -55& (f)\\
PA$\_$opt($\circ$) & 10$\pm$ 5 & (b)\\
Inclination$\_$opt($\circ$)&67$\pm$ 3&(b)\\
Total HI mass (M$_{\odot}$)&1.34$\times10^{8}$& (f)\\
Dynamical mass (M$_{\odot}$) & 3.2$\times 10^{9}$& (f)\\
\hline     \\
\multicolumn{3}{@{} p{8.8 cm} @{}}{\footnotesize{\textbf{Notes.} Ref\,(a) \citet{1991Sci...254.1667D}; (b) \citet{1998ARA&A..36..435M};
(c) \citet{2000ApJ...537L..95D}; (d) \citet{2003MNRAS.340...12W}; (e) \citet{1989MNRAS.240..563S}; (f) \citet{2006AJ....131..363D}; (g) \citet{2012MNRAS.421.2998F}}}
\label{coords_table}
 
\end{tabular}   

\end{minipage}
\end{table}  

\section{KAT-7 Observations and Data Reduction}
A mosaic of three pointings was observed for NGC 6822 in the 21cm line with the seven dish KAT-7 array, a MeerKAT engineering test bed \citep{2013AJ....146...48C}. The array is compact with baselines ranging from 26 to 185m. The compact baselines and the low receiver temperature (T$_{\text{sys}}$ $\simeq 26 $K) \citep{2016MNRAS.460.1664F} of KAT-7 allow us to detect low surface brightness extended emission 
\citep{2015MNRAS.450.3935L,2017MNRAS.464..957H}.
The observations were done between August and December 2014. The observations were done in 20, 5-8 hr segments for the total integration time of $\sim$ 151 hrs, including calibration. 17 sessions were observed using a full 7 antenna array while 3 sessions were observed with 5 or 6 antennas depending on the availability of antennas present. The observation parameters are listed in Table 2. The c16n13M4K correlator mode set up was used, that is we measured two linear polarizations in 4096 channels over a bandwidth of 12.5 MHz allowing for the spectral resolution of 0.64 kms$^{-1}$, centered at 1420.7 MHz. 

The calibration steps followed the standard techniques of the CASA software package version 4.3.0 \citep{2007ASPC..376..127M}. PKS 1934-638 and PKS 1938-155 were used as the primary and secondary calibrators respectively. Data were loaded into CASA and inspected for antenna failures and Radio Frequency Interference (RFI) using the CASA task VIEWER. An automated flagging was used to discard data resulting from shadowing and low elevation $\leqslant$ 20 deg. 10 $\%$ of the channels on both sides of the band were flagged out to get rid of the bandpass roll off. We flagged out the entire antenna 7 on 4 observing runs due to severely bad data caused by a warm cryo. Flagging out antenna 7 from these observations meant that we lost out on most of the long baseline observations. This resulted in 13 observing runs having data with all the 7 antennas present during observations and not 17 as initially stated before flagging. The data was Hanning smoothed by averaging 4 channels giving the velocity resolution of 2.56 kms$^{-1}$. We then proceeded with the absolute flux calibration as well as the bandpass and gain calibration. The quality of the calibration solutions were first checked by inspecting the phase and amplitude calibration tables, and applying the solutions to the primary and secondary calibrators respectively. The corrected visibilities were examined in terms of amplitude and phase as a function of time and frequency as well as amplitude and phase as a function of uv distance for both calibrators. The examined calibration solutions were then applied to the target source for each observing run. 

The calibrated target visibilities for the three mosaic pointings were separated from the calibrator sources and different observing runs were then combined together using the CASA task CONCAT. To ensure the angular resolution of our observation was not compromised due to the missing long baseline antennas on some of the observing runs, we decided to create two separate files. The first file was created by combining 13 observing files, each having all the 7 antennas present during observation and after flagging while the second file was created by combining all the 20 observing sessions. The task CVEL was then used to produce data files with proper velocity coordinates as well as correct for the frequencies from scan to scan. Detailed used of CVEL on KAT-7 data has been described in \citep{2013AJ....146...48C}. Velocities from -800 to -200 and 200 to 600 kms$^{-1}$ were chosen as representative of the continuum and the task UVCONTSUB was used to subtract the continuum from the spectral data. This was carefully done by taking note of the Galactic HI contamination.
\subsection{Imaging}
Imaging was done using the task CLEAN by selecting the mosaic and multi-scale clean option. Multi-scale clean allows us to retain the extended low level emission by modeling the sky brightness as the summation of components of emission having different size scales. A dirty channel map was produced for each data set produced from the CONCAT task to compare the resolution and sensitivity. Inspecting the two maps showed that we do not lose resolution by combining observations runs with missing long baseline antennas, however there was no significant change in the signal to noise between the images produced by the two data sets. Considering the data volume and processing time, we decided to work with the data set produced by combining the observation runs with all 7 antennas present during observations and after flagging.

A mask was first created by using the interactive clean option and selecting regions of galaxy emission manually. The rms was then established by measuring the average noise in four line free channels. The final cube was imaged in a non interactive mode using the mask created and cleaning down to a 1.5 rms threshold. The 1.5 rms threshold was selected after investigating different flux cutoff. Thresholds less than 1.5 rms showed clean artifacts of over cleaning while the threshold greater than 1.5 rms was not sufficient as we were unable to retain all the galaxy flux. Three final cubes were produced by applying the natural (na), uniform (un), and the neutral (robust=0) weighting to the data. We decided to work with the robust=0 cube after comparing the sensitivity and the resolutions obtained from the three weighting options. With our angular resolution (synthesized beam) of $\sim$ 3.5$^{\prime}$, the cubes were imaged using the pixel size of 30$^{\prime \prime}$ by 30$^{\prime \prime}$ to ensure a well resolved central lobe ($\sim$ 6 pixels in beam). 

We noticed the presence of horizontal stripes in the image plane of all our cubes. These artifacts have been attributed to the presence of radio frequency interference (RFI) generated internally around $|u| \leqslant$ 25 $\lambda$ \citep{2016A&A...587L...3C,2015MNRAS.452.1617H}. We created three cubes to test the cutoff limits for the artifacts in the cube. The first cube was produced after flagging out visibilities with $|u| \leqslant 10  \lambda$, the second cube was produced after flagging out visibilities with $|u| \leqslant 15 \lambda$ and lastly the third cube was produced after flagging out visibilities with $|u| \leqslant 25 \lambda$. The artifacts were seen to disappear in all the three cubes. This allowed us to select the cube created by flagging data ($|u| \leqslant 10 \lambda$), the advantage being less percentage of data was flagged in this case enabling us to reserve the much needed short spacing data. We uncovered further artifacts in form of broader horizontal diagonal lines parallel to the major axis of NGC 6822. These artifacts were removed by getting rid of the affected baseline between antennas 4 and 5 for all observations. After getting rid of these artifacts, the noise level in the cube reduced by a factor of $\sim$ 2 allowing us to reach the expected theoretical noise in line free channels.

Given the radial velocity of NGC 6822, the H\textsc{i} data cube is heavily affected by foreground Galactic emission in the range -18 kms$^{-1}$ to 24 kms$^{-1}$. The Galactic emission between the velocity range 8 kms$^{-1}$ to 24 kms$^{-1}$ is spatially separated from the galaxy's emission and could thus be subtracted with ease. However, Galactic emission between -18 kms$^{-1}$ to 7 kms$^{-1}$ is much closer to the galaxy's emission and was blanked using the following criteria: in all channels affected by Galactic emission, we isolated emission 3 times the rms noise in line free channels (3$\sigma$, where $\sigma$ = 3.7 mJy/beam). Plotting the H\textsc{i} profiles at each pixel position revealed that the Galactic H\textsc{i} was mainly visible in 3 consecutive channels (i.e the Galactic H\textsc{i} has narrow velocity width as compared to the galaxy's emission). Using this information, we were able to consider emission detected in more than 3 consecutive channels as real galaxy emission. Any emission spanning across $\leqslant$ 3 channels was blanked. In a few positions, the Galactic H\textsc{i} was seen across 5 consecutive channels. In these cases, we considered emission detected in more than 5 consecutive channels as real galaxy emission. The above process was repeated for the entire velocity range to check how much galaxy flux could be lost when Galactic emission was considered to span across 3 channels. A flux loss of $\sim$ 0.3$\%$ was calculated. This shows that the above process does not interfere with the galaxy emission, but is able to subtract well the Galactic HI from the galaxy. 

The final cube was created by removing noise pixels in each channel. The original cube was first smoothed to twice the original resolution. All the pixels below the 3$\sigma$ of the smoothed image were blanked out. The resulting cube still had some imperfections in regions with very bright emission. These were blanked by hand. The final masked cube was later used to derive the H\textsc{i} global profile, the intensity H\textsc{i} maps, the H\textsc{i} velocity maps and the H\textsc{i} dispersion maps.  

\begin{table}
\scriptsize
\caption{\small Parameters of the KAT-7 Observations.}
\begin{minipage}{\textwidth}
\begin{tabular}{l@{\hspace{1.2cm}}c@{\hspace{1.2cm}}}   
\hline

Parameter & Value \\
                         %&(sec)       &(Mpc) & (mag) & ($L_{\odot}$) &($\rm M_{\odot}yr^{-1}$)  \\
                %~~~~~~(1)    &   (2)        &  (3)   \\         
   
\hline \hline  
%\multicolumn{6}{@{} p{8.5 cm} @{}}{\footnotesize{\hspace{3cm} VLT/NACO DATA}}\\\\
Start of observations & August 2014 \\
End of observations & December 2014\\
Total integration per pointing & $\sim$ 35 hrs\\
FWHM of primary beam & 1.27$^{\circ}$\\
Total bandwidth & 12.5 MHz \\
Channel width (4 $\times$ 0.64 kms$^{-1}$) & 2.56 kms$^{-1}$\\
Number of channels (4096/4)&1024 \\
Map gridding &  30$^{\prime \prime}$ by 30$^{\prime \prime}$\\
Map size & 256 by 256\\
Flux/bandpass calibrator& 1934-638\\
Phase calibrator & 1938-155 \\
\hline    
\multicolumn{2}{@{} p{8.5 cm} @{}}{\footnotesize{\hspace{1cm} Robust = 0 weighting function}}\\\\
FWHM of synthesized beam & 209$^{\prime \prime} \times$ 193$^{\prime \prime}$  \\
RMS noise & 3.7 mJy/beam\\
Column density limit&\\
(3$\sigma$ over 16 kms$^{-1}$) & 1.0$\times 10^{19}$ cm$^{-2}$ \\
\hline
%\multicolumn{6}{@{} p{8.5 cm} @{}}{\footnotesize{\textbf{Notes.} Col\,(1): {\tt IRAS} survey name \textcolor{red}{with the old NACO data marked with an asterisk}; (2): Total exposure time; (3): Luminosity distance from NED Database; (4):  $K_S$-band absolute magnitude of the brightest cluster; (5): Galaxy IR luminosity from \citet{2003AJ....126.1607S}, any value marked by $\dagger$ is estimated by using the method described in $\S$\,\ref{sec_relation}; (6): SFR derived from Eq.\,\ref{Kennicut}}}
\label{coords_table}
 
\end{tabular}   

\end{minipage}
\end{table}  
%\begin{figure}
%\centering
%\resizebox{1.0\hsize}{!}{\rotatebox{0}{\includegraphics{fig1.pdf}}}
%\caption{\small  An empirical relation between the $K$-band magnitude of the brightest cluster and the SFR of the galaxy. The dashed line shows a weighted linear fit to all the data, including the three most distant targets at $D_L > 150\,\rm Mpc$ shown as open squares. The solid line fits the $D_L \leq 150\,\rm Mpc$ targets labeled as circles;  those at $D_L \leq 100\,\rm Mpc$  are black and those at $100 < D_L \leq 150\,\rm Mpc$ are grey.}
%\label{relation}
%\end{figure}
\section{H\textsc{i} Distribution}
The global H\textsc{i} emission profile of NGC 6822 is shown in Figure 1. The profile was derived from the primary beam corrected masked channel maps. We find H\textsc{i} emission between -118 to 7 kms$^{-1}$. This is similar to the range of velocities detected by \citet{2000ApJ...537L..95D}. A mid-point velocity of -55 $\pm$ 2 kms$^{-1}$ is found. This is identical to the value found by \citet{2006AJ....131..343D}. The profile widths at 20$\%$ and 50$\%$ levels are 105 $\pm$ 0.6 kms$^{-1}$ and 90 $\pm$ 2 kms$^{-1}$. We integrated the flux over the global profile and measured the total integrated flux of 2440 $\pm$ 20 Jy.kms$^{-1}$. Adopting a distance to NGC 6822 of 0.48 Mpc \citep{2012MNRAS.421.2998F}, we find the total H\textsc{i} mass of M$_{\text{H\textsc{i}}}$ =  (1.3 $\pm$ 0.1) $\times 10 ^{8}$ M$_{\odot}$. This is 23 $\%$ more than the ATCA measurements \citep{2006AJ....131..343D} (M$_{\text{\text{H\textsc{i}}}}$ = 1.0 $\times$ 10$^{8}$M$_{\odot}$) and very close to the Parkes single dish measurement of \citep{2006AJ....131..343D} (M$_{\text{\text{H\textsc{i}}}}$ = 1.2 $\times 10^{8}$ M$_{\odot}$), all were corrected to the distance of 0.48 Mpc \citep{2012MNRAS.421.2998F}. 
\begin{figure}
  \includegraphics[width = \columnwidth]{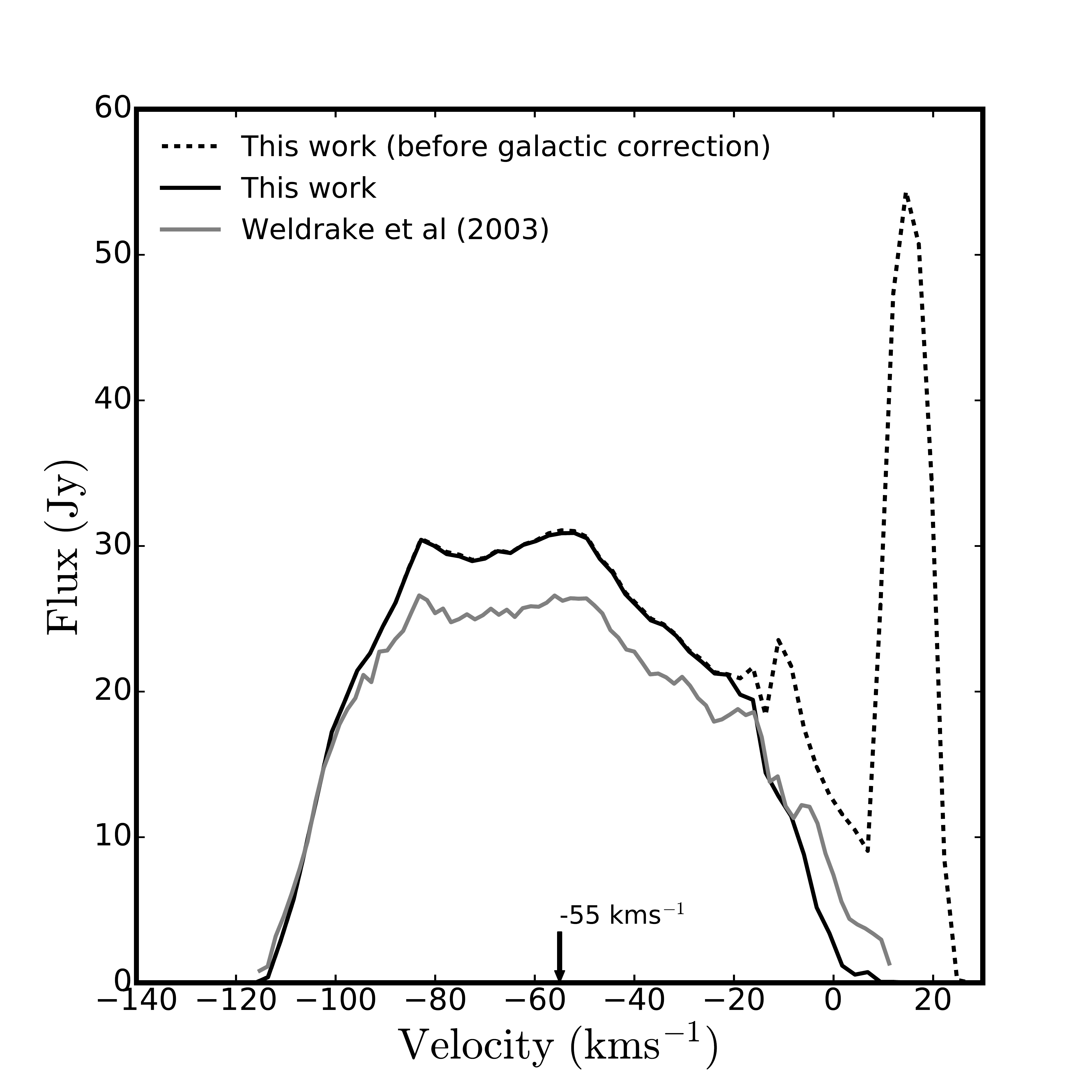}
  \caption{H\textsc{i} emission for NGC 6822 with the dash line showing the profile not corrected for Galactic emission. The black solid line shows the KAT-7 profile after the Galactic emission has been subtracted while the grey solid line shows the ATCA profile corrected for Galactic HI. The systemic velocity of the profile is shown at -55 kms$^{-1}$.} %The inset histogram show the distribution of the difference whereby the dash vertical lines indicate $|\Delta{z}|\,/(\,1\,+\,z_{spec})\,=\,0.2$. 
 % The mean $(\mu)$, and standard deviation $(\sigma)$ of the distribution are indicated in the top right of the panel}
  \label{fig_speczvsphotz} 
\end{figure}

Figure 2 shows the H\textsc{i} column density map superposed on the optical DSS image. We integrated the H\textsc{i} column densities from -118 to 7 kms$^{-1}$ in the KAT-7 primary beam corrected cube. The lowest contour in our column density map is 1$\times 10^{19}$ cm$^{-2}$. This is consistent with the 3$\sigma$ noise limit calculated over 16 kms$^{-1}$. Our derived column density limit is significantly better than the 3$\sigma$ sensitivity of 4$\times 10^{20}$ cm$^{-2}$ calculated over 16 kms$^{-1}$ of \citet{2000ApJ...537L..95D}. The H\textsc{i} emission is seen to be much larger than the optical body, which is typical of irregular galaxies. An H\textsc{i} diameter of $\sim$ 1.2$^{\circ}$ is measured along the major axis at the lowest contour. This is in agreement with the Parkes single dish measurement \citep{2000ApJ...537L..95D}. 

The H\textsc{i} column density grey scale map of Figure 3 shows structures present in the H\textsc{i} disk. The H\textsc{i} hole as well as the arm in the South East previously reported by \citet{2000ApJ...537L..95D} are clearly visible in the KAT-7 observations. The velocity field map of NGC 6822, shown at the top of Figure 4, is seen to have symmetric and undisturbed velocities similar to most low mass, late-type galaxies. This will be used to derive the kinematical properties of the galaxy using the tilted ring model explained in Section 4. At the bottom of Figure 4, we can see that the velocity dispersion varies from 10 kms$^{-1}$ in the center to $\sim$ 4 kms$^{-1}$ at the edge of the disk.

\begin{figure}
  \includegraphics[width = \columnwidth]{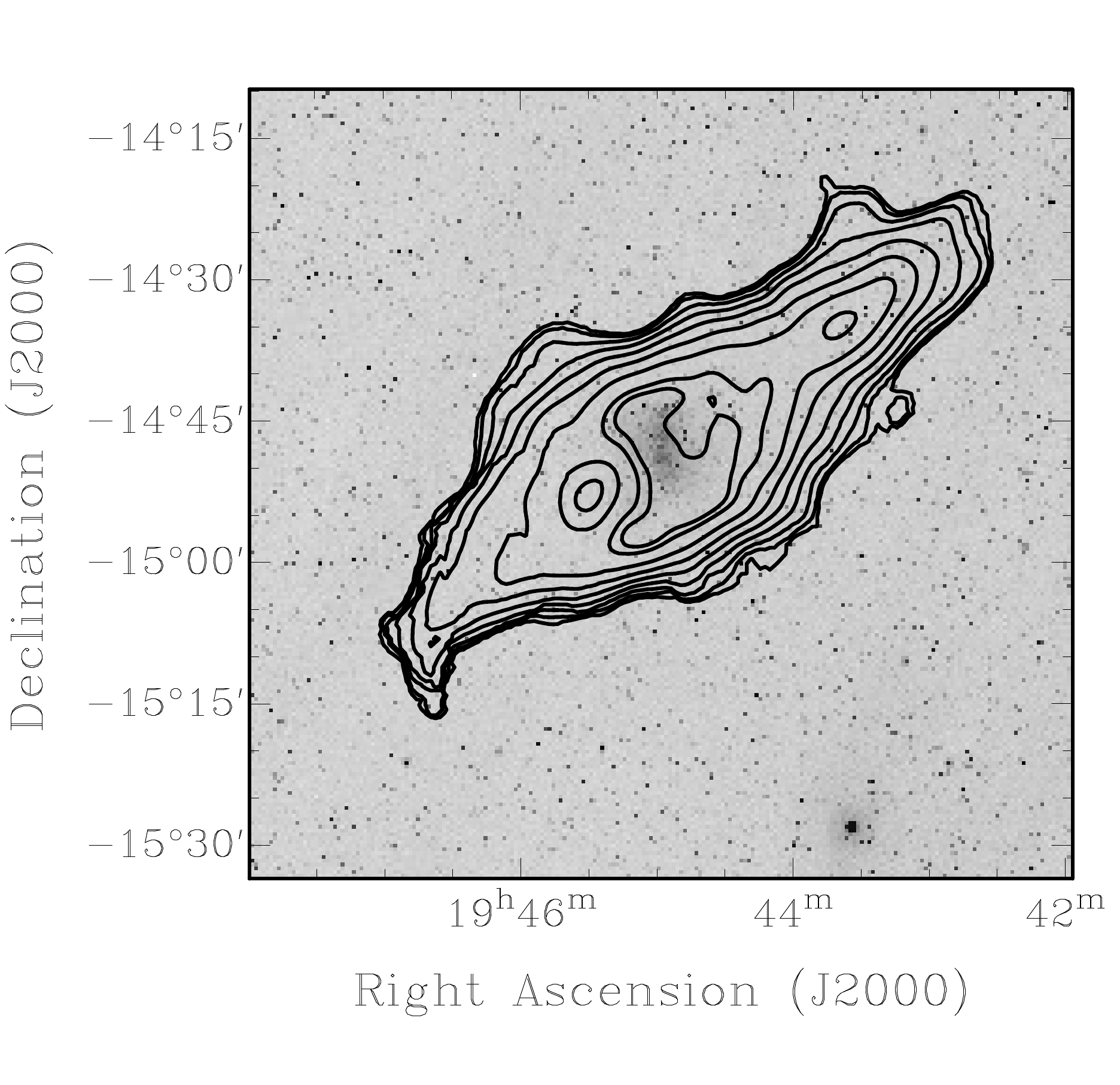}
  \caption{KAT-7 total column density collapsed map of NGC 6822 overlaid on an optical DSS image. The column density contours are 0.01, 0.02, 0.04, 0.08, 0.16, 0.32, 0.64, 1.28, and 1.60 $\times 10^{21}$ cm$^{-2}$. The lowest contour is at a 3$\sigma$ level.}
   %The inset histogram show the distribution of the difference whereby the dash vertical lines indicate $|\Delta{z}|\,/(\,1\,+\,z_{spec})\,=\,0.2$. 
 % The mean $(\mu)$, and standard deviation $(\sigma)$ of the distribution are indicated in the top right of the panel}
  \label{fig_speczvsphotz} 
\end{figure}

\begin{figure}
  \includegraphics[width = \columnwidth]{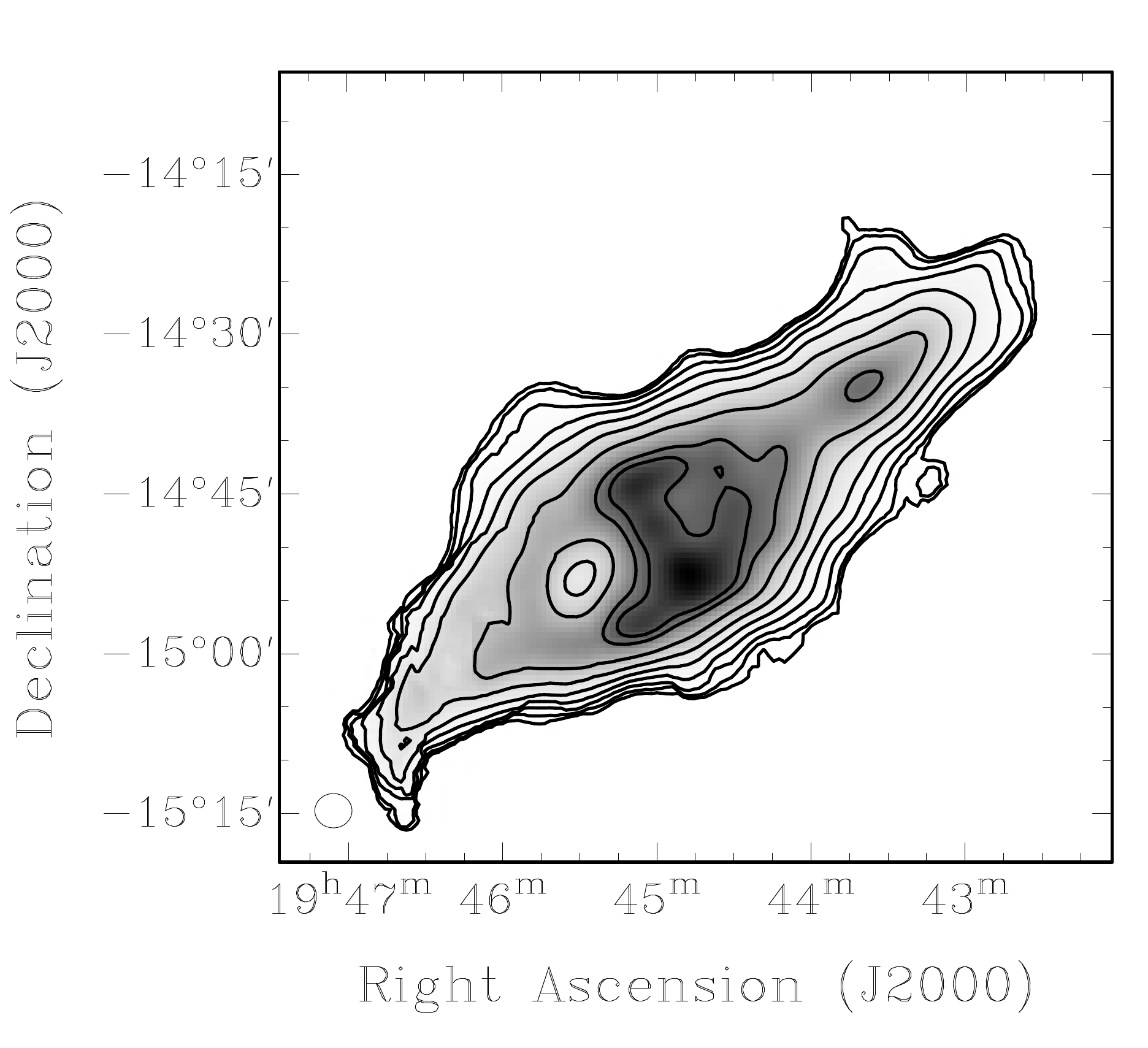}
  \caption{Integrated H\textsc{i} column density map. The greyscale levels run from 1$\times 10^{19}$ cm$^{-2}$(white) to 2.52 $\times 10^{21}$ cm$^{-2}$ (black). The column density contours are 0.01, 0.02, 0.04, 0.08, 0.16, 0.32, 0.64, 1.28, and 1.60 $\times 10^{21}$ cm$^{-2}$. The beam is shown in the bottom left corner. An H\textsc{i} hole is clearly seen SE of the bright central parts.}
   %The inset histogram show the distribution of the difference whereby the dash vertical lines indicate $|\Delta{z}|\,/(\,1\,+\,z_{spec})\,=\,0.2$. 
 % The mean $(\mu)$, and standard deviation $(\sigma)$ of the distribution are indicated in the top right of the panel}
  \label{fig_speczvsphotz} 
\end{figure}

\begin{figure}
    \centering
    \begin{subfigure}[b]{0.45\textwidth}
        \includegraphics[width=\textwidth]{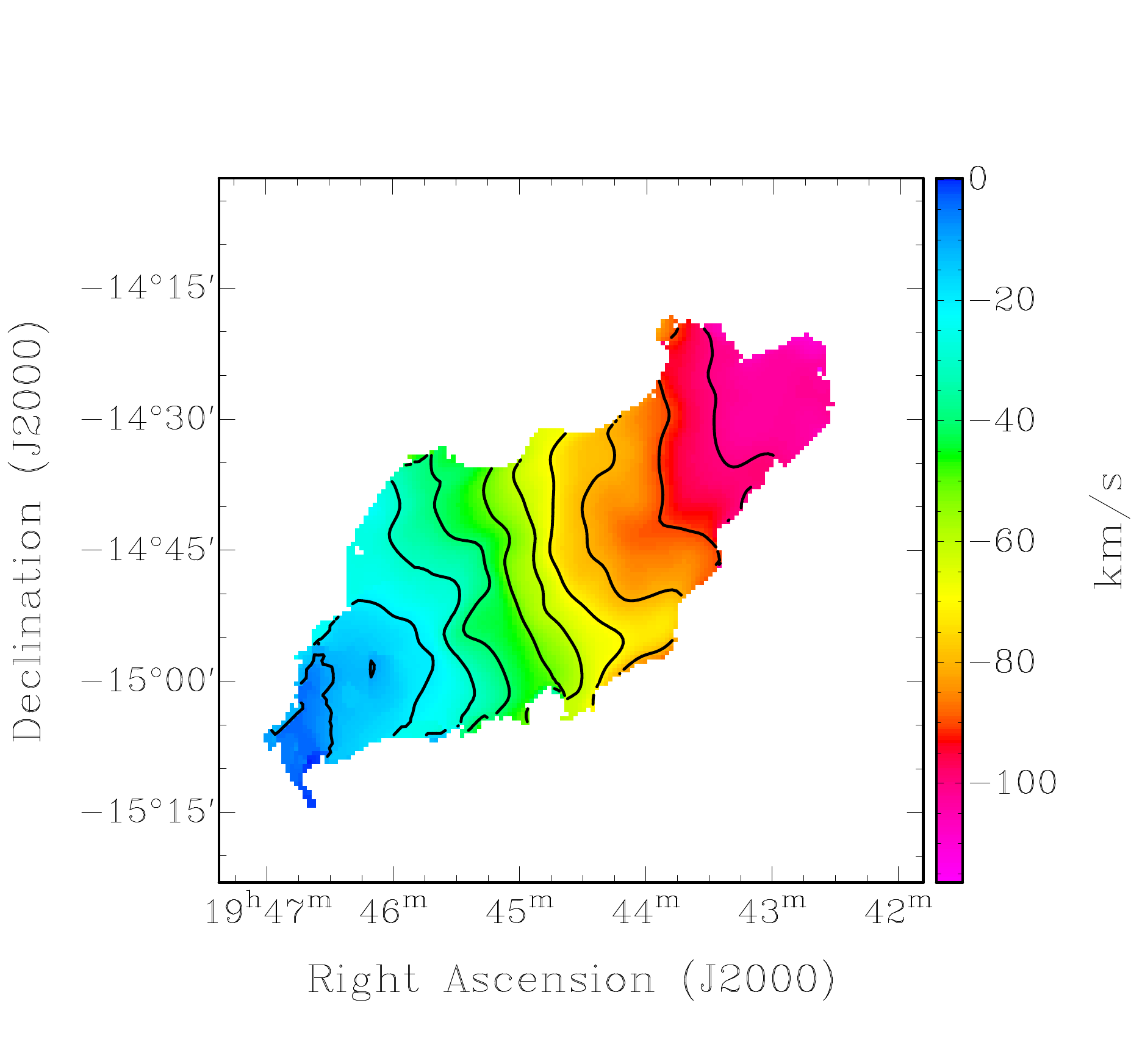}
        \caption{Velocity field map}
        \label{fig:gull}
    \end{subfigure}
    ~ %add desired spacing between images, e. g. ~, \quad, \qquad, \hfill etc. 
        ~ %add desired spacing between images, e. g. ~, \quad, \qquad, \hfill etc. 
    \begin{subfigure}[b]{0.45\textwidth}
        \includegraphics[width=\textwidth]{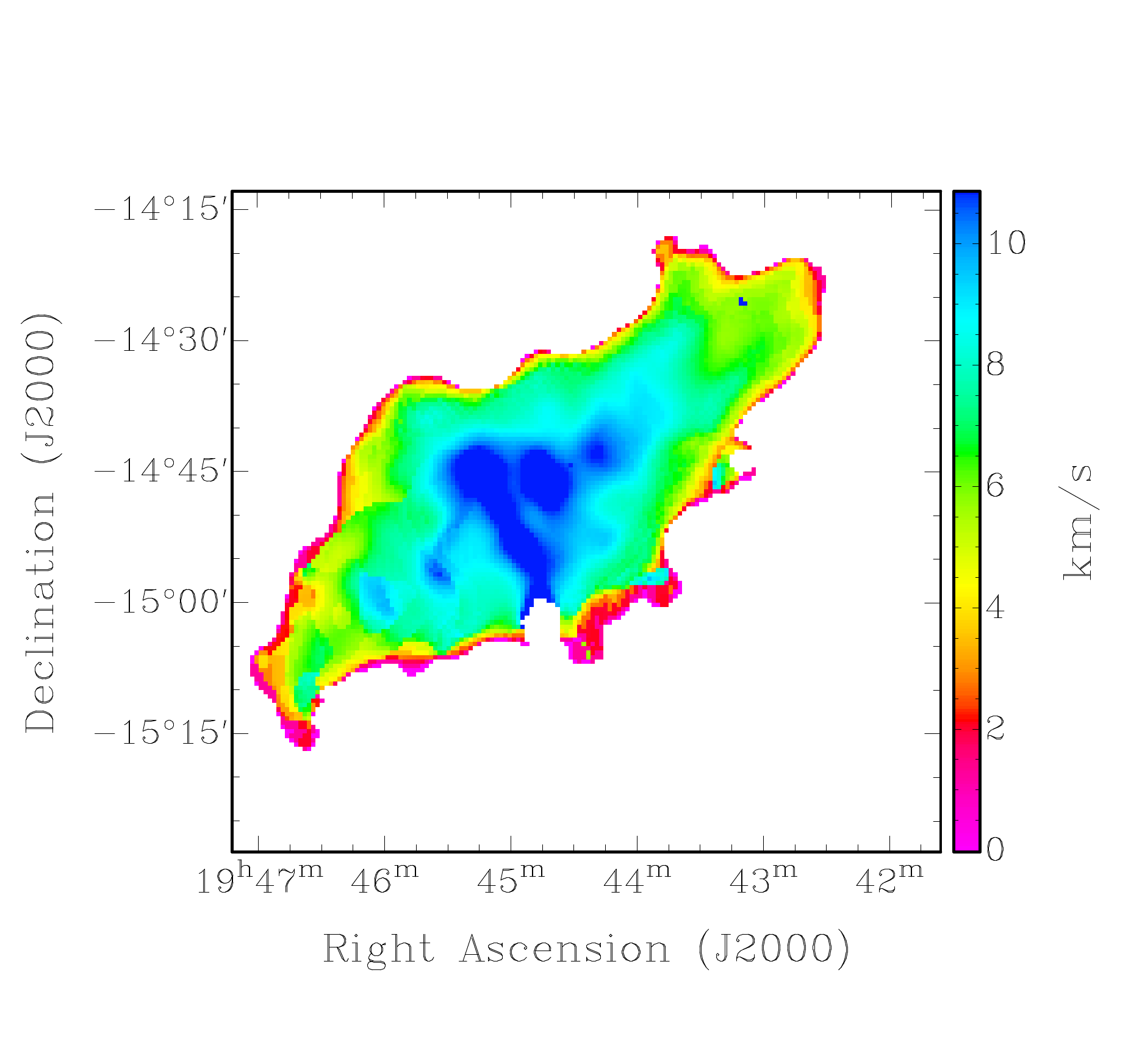}
        \caption{Dispersion map}
        \label{fig:mouse}
    \end{subfigure}
    \caption{First (top) and second (bottom) moment maps of NGC 6822 from the KAT-7 data cube. The top shows the observed velocity field. The contours are -100, -90, -80, -70, -60, -50, -40, -30, -20, and -10 kms$^{-1}$. The bottom shows the observed velocity dispersion map.}
    \label{}
\end{figure}
%\begin{figure}
%  \includegraphics[width = \columnwidth]{finvelocity.eps}
%  \caption{Velocity field map of NGC6822. The velocity contours are -100, -90, -80, -70, -60, -50, -40, -30, -20, -10, 0, and 10 kms$^{-1}$}}
   %The inset histogram show the distribution of the difference whereby the dash vertical lines indicate $|\Delta{z}|\,/(\,1\,+\,z_{spec})\,=\,0.2$. 
 % The mean $(\mu)$, and standard deviation $(\sigma)$ of the distribution are indicated in the top right of the panel}
%  \label{fig_speczvsphotz} 
%\end{figure}
Figure 5 shows the azimuthally averaged radial H\textsc{i} density profile of NGC 6822. This is compared to the radial profile derived from the previous ATCA observation, smoothed to the KAT-7 resolution \citep{2003MNRAS.340...12W}.The profile was derived using the GIPSY task ELLINT by applying the tilted ring kinematical parameters described in Section 4. The values for the derived HI surface densities are given in Table 3.

\begin{figure}
  \includegraphics[width = \columnwidth]{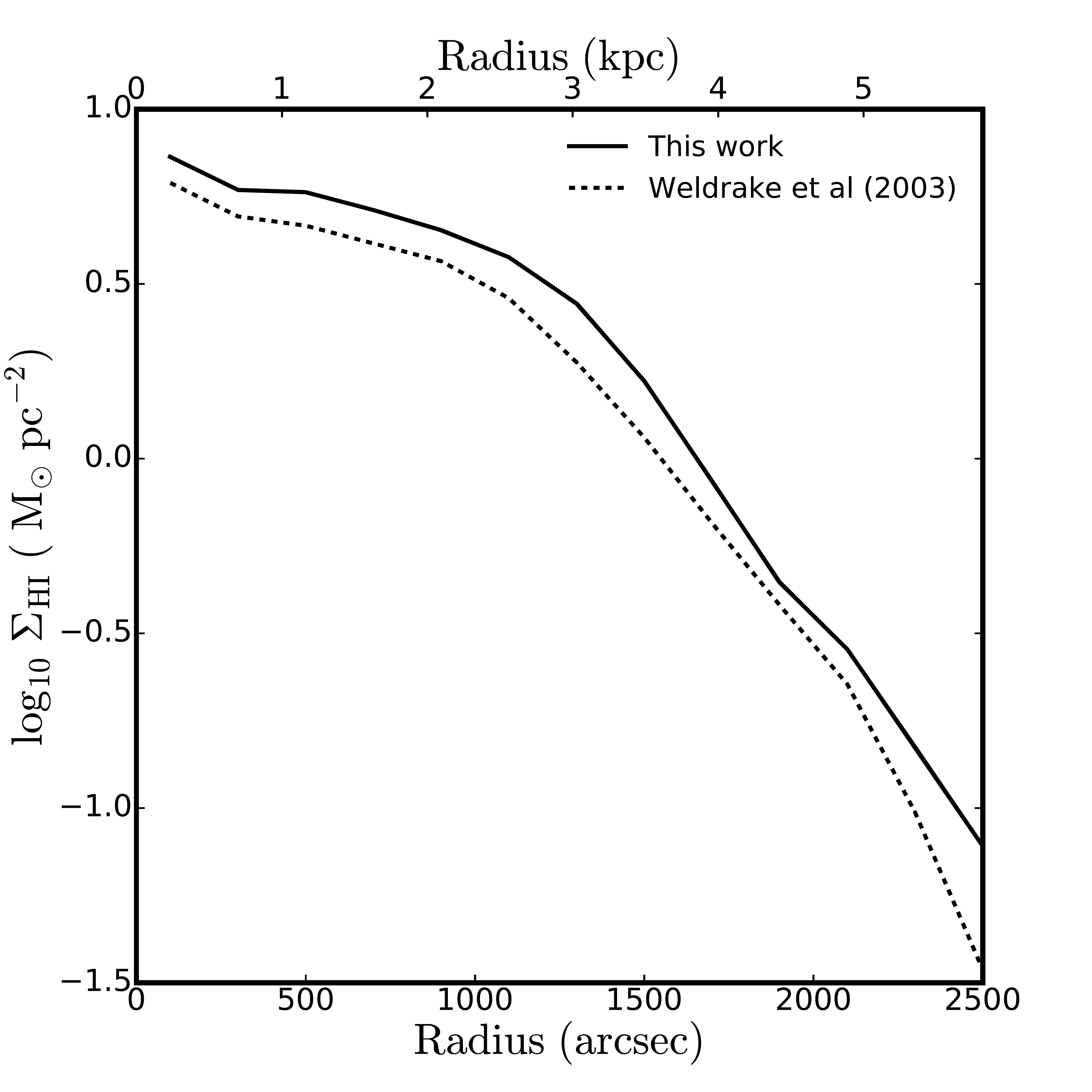}
  \caption{Comparison of the ATCA \citep{2003MNRAS.340...12W} and the KAT-7 radial profiles. The profiles have been corrected for helium and heavy  metals by multiplying the surface densities by 1.4.}
   %The inset histogram show the distribution of the difference whereby the dash vertical lines indicate $|\Delta{z}|\,/(\,1\,+\,z_{spec})\,=\,0.2$. 
 % The mean $(\mu)$, and standard deviation $(\sigma)$ of the distribution are indicated in the top right of the panel}
  \label{fig_speczvsphotz} 
\end{figure}

\section{HI KINEMATICS}
\subsection{Tilted ring model}
 A tilted ring model \citep{1974ApJ...193..309R} is based on the assumption that the velocities we observe are mainly circular rotational velocities and that the disk galaxy can be described by a set of concentric rings. Each ring has a set of defining kinematical parameters; central co-ordinates (x$_{c}$, y$_{c}$), a systemic velocity V$_{\text{sys}}$, an inclination $i$, a position angle (P.A), and the rotation velocity, V$_{\text{rot}}$. The line of sight velocity at any (x,y) position in a ring with radius R is given by
 \begin{equation}
 V(x,y) = V_{\text{sys}} + V_{\text{rot}} \sin(i)\cos(\theta),
 \end{equation}
 where $\theta$ is the position angle with respect to the receding major axis measured in the plane of the galaxy, $\theta$ is related to the actual
 P.A. in the plane of the sky by
\begin{equation}
 \cos(\theta) = \frac{-(x-x_{0}) \sin(P.A.) + (y -y_{0}) \cos(P.A.)}{R}.
 \end{equation}
\begin{equation}
  \sin(\theta) = \frac{-(x-x_{0}) \cos(P.A.) + (y -y_{0}) \cos(P.A.)}{R \cos(i)}. 
\end{equation}
\subsection{Deriving the rotation curve}
We derived the rotation curve of NGC 6822 by applying the tilted ring model to the velocity field using the GIPSY\citep{1992ASPC...25..131V} task ROTCUR \citep{1989A&A...223...47B}. We determined parameters for each ring independently by using a ring width and spacing between rings corresponding to the beam width of $\sim$ 200 arcsec. Our final ring radius was 2500 arcsec because beyond this radius the S/N dropped drastically. The parameters that were fitted are: the dynamical center (x$_{c}$, y$_{c}$), the systemic velocity V$_{\text{sys}}$, the position angle PA, the inclination $i$, and the rotation velocity V$_{\text{rot}}$. We set V$_{\text{exp}}$, the expansion velocity, to zero. At first, we ran the tilted ring model with all the parameters free out to the edge of the optical disk (R$_{25}\sim$ 1.1 kpc \citep{1991Sci...254.1667D}). The V$_{\text{sys}}$ and the rotation center are expected to be the same for all rings, at least within the optical disk. From this, the average values of the V$_{\text{sys}}$ and the (x$_{c}$, y$_{c}$) were fixed. The fixed systemic velocity and the dynamical center were used to run the tilted ring model out the edge of the H\textsc{i} disk. We iterated by fixing either the position angle or the inclination or both to derive the best fit model. As the minor axis provides little information regarding the rotation curve, an exclusion angle of 15$^{\circ}$ was used to give more weight to the points around the major axis. The data were also weighted by applying a $|cos(\theta)|$ weighting, where $\theta$ is the azimuthal angle in the plane of the galaxy from the major axis. Once we produced the best fit rotation curve for both sides, we also derived the curves for the approaching and receding sides to see possible departure from axisymmetry. The errors on V$_{\text{rot}}$ were derived using Equation 4 \citep{2013AJ....146...48C}.
\begin{equation}
\Delta V = \sqrt{\sigma^{2}(V) + \Bigg(\frac{|V_{\text{app}} - V_{\text{rec}}|}{2}\Bigg)^2},
\end{equation}
where $\sigma_{V}$ is the intrinsic dispersion inside the rings and V$_{\text{app}}$ and V$_{\text{rec}}$ are respectively the approaching and receding side velocities.

Equation (4) will give larger errors than what you sometimes see in the literature where people only use the intrinsic dispersion in the rings as their errors. However, since at the mass model stage, we will fit this rotation curve with axisymmetric components, we think the differences between both sides of the galaxy are more representative of the true uncertainties in those models. 

The velocity dispersions in NGC 6822 are large enough to contribute to the gravitational support of the galaxy. In order to obtain more reliable rotation velocities for the galaxy, we correct for the pressure gradient in the gas using the asymmetric drift correction. Following the method described in \citet{2000AJ....120.3027C}, we correct for the asymmetric drift as follows:
\begin{equation}
V_{c}^{2} = V_{0}^{2} - 2\sigma \frac{\delta \sigma}{\delta \ln R } - \sigma^{2} \frac{\delta \ln \Sigma}{\delta \ln R},
\end{equation}
where $V_{c}$ is the corrected velocity, V$_{0}$ is the observed one, $\sigma$ is the velocity dispersion and $\Sigma$ is the gas density. Table 3 shows the parameters used to correct for the asymmetric drift together with the corrected velocities. 
\subsection{KAT-7 rotation curve and position velocity diagram}
The resulting rotation curves (approaching, receding, both sides) of NGC 6822 derived from the velocity field are shown in Figure 6. We adopted the model with constant inclination and varying position angle due to the following reasons: 1) it is difficult to constrain the varying inclination in the inner region of the galaxy due to the presence of the bar major axis which appears to be perpendicular to the major axis of the disk as shown in Figure 2, and 2) deriving surface densities using varying inclinations derived from the tilted ring parameters show that the inner regions of the galaxy have low H\textsc{i} column densities as compared to the outer regions of the galaxy which is not consistent with the data as shown in Figure 3. We derive the dynamical center of the galaxy at $\alpha, \delta$ (J2000) = 19h44m58.0s, -14d48m11.9s, and its systemic heliocentric velocity, $V_{\text{sys}}$ = -55 $\pm$ 3 kms$^{-1}$. This is similar to the global profile value. The dynamical center agrees within 1 arcsec, with the optical center $\alpha, \delta$ (J2000) = 19h44m57.9s and DEC -14d48m11.0s \citep{1991Sci...254.1667D}. We derive the mean P.A. = 118$^{\circ}$ and $i$ = 66$^{\circ}$. Between 500 $\sim$ 900 arcsec, the difference between the approaching and receding sides become more prominent for $i$ and PA but, curiously, the smallest for $V_{\text{rot}}$. This correspond to the edge of the inner disk. At velocities $>$ 2000 arcsec, the difference between the approaching and receding side can be attributed to the residual Galactic HI contamination. In this case, the receding side which is affected by the Galactic H\textsc{i} is seen to have higher rotation velocity values as compared to the approaching side. 

Our derived kinematics of NGC 6822 are in agreement with the literature. \citet{2003MNRAS.340...12W} derived the V$_{\text{sys}}$ = 54.4 kms$^{-1}$, mean P.A. = 118$^{\circ}$, and the mean $i$ = 63$^{\circ}$ of NGC 6822 from their tilted ring results. The KAT-7 RC agree very well with the ATCA data, see Figure 7 while allowing us to extend the rotation data $\sim$ 1 kpc further out. The consistency between the derived kinematical parameters from both observations show that the larger extent of the  KAT-7 RC comes from the increased sensitivity which allows us to detect the low column density gas in the outer radii.  

The best way to check that our derived rotation curve best represents the kinematics of the disk is to overlay the rotation curve on a position-velocity (PV) diagram obtained along the major axis, as shown in Figure 8. The rotation curve shows a good representation of the dynamics of the galaxy. 

We have also compared the observed velocity field with the model velocity field constructed from the tilted ring fits for both, approaching and receding sides. These are shown in Figure 9. The residual maps derived from the model velocity maps for each side show similar residual pattern with no large residuals seen. Most of the residuals are $< \pm$ 10 kms$^{-1}$. An increase of the residuals at the H\textsc{i} hole position is clearly seen. 
\begin{figure*}
\centering
\resizebox{1.0\hsize}{!}{\rotatebox{0}{\includegraphics{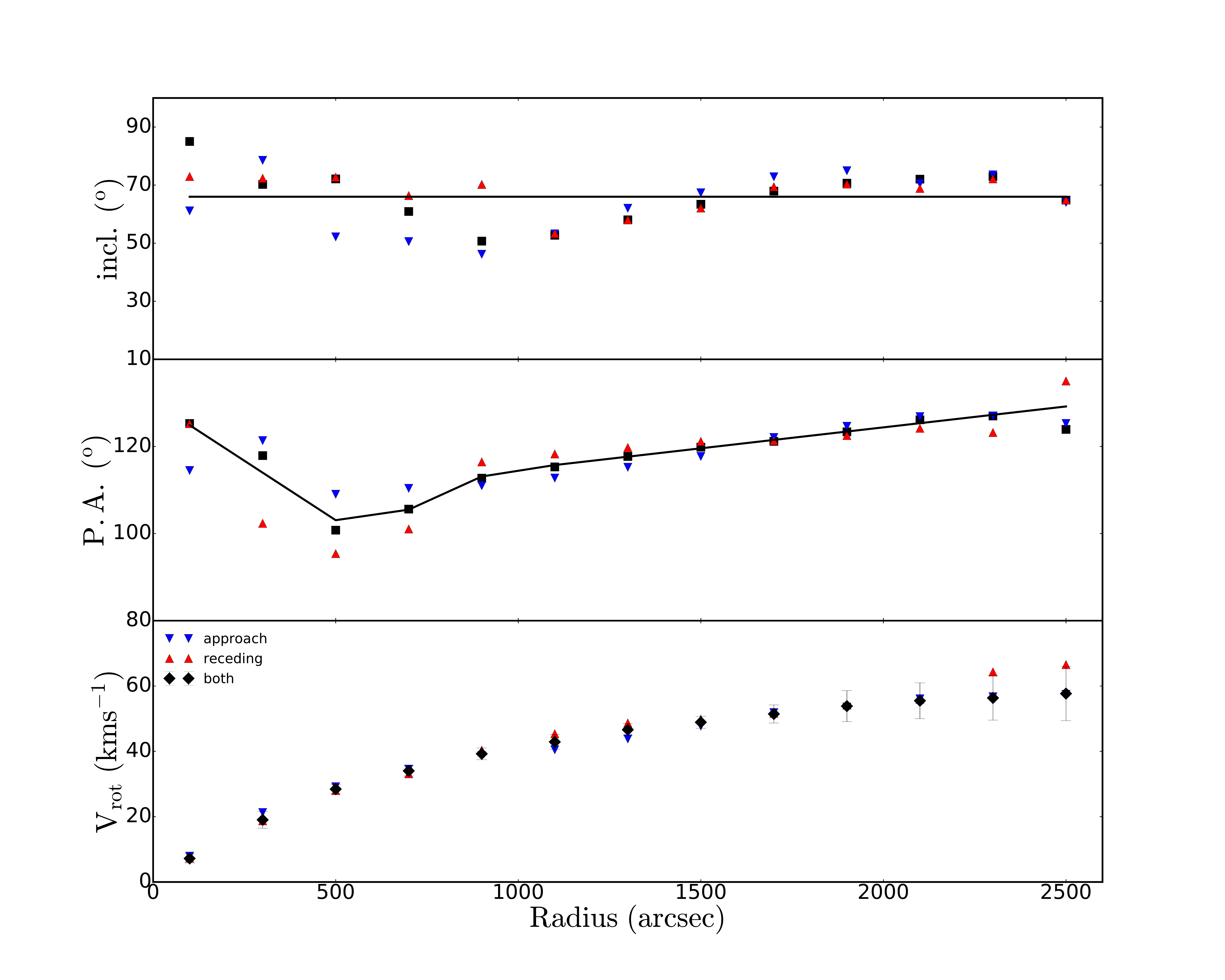}}}
\caption{Results of the tilted ring fits for NGC 6822. For the middle and top panels, the black dotted lines show the behavior of the PA and inclination as free parameters while the black dashed lines show the behavior of the PA and inclination fixed to the model used to derive the final rotation curve. In this case the PA is varying while the inclination is fixed to the mean value. For the bottom panel, the blue triangles represent the curve for the approaching side while the red triangles represents the curve for the receding side.}
   %The inset histogram show the distribution of the difference whereby the dash vertical lines indicate $|\Delta{z}|\,/(\,1\,+\,z_{spec})\,=\,0.2$. 
 % The mean $(\mu)$, and standard deviation $(\sigma)$ of the distribution are indicated in the top right of the panel}
  \label{fig_speczvsphotz} 
\end{figure*}

\begin{figure}
  \includegraphics[width = \columnwidth]{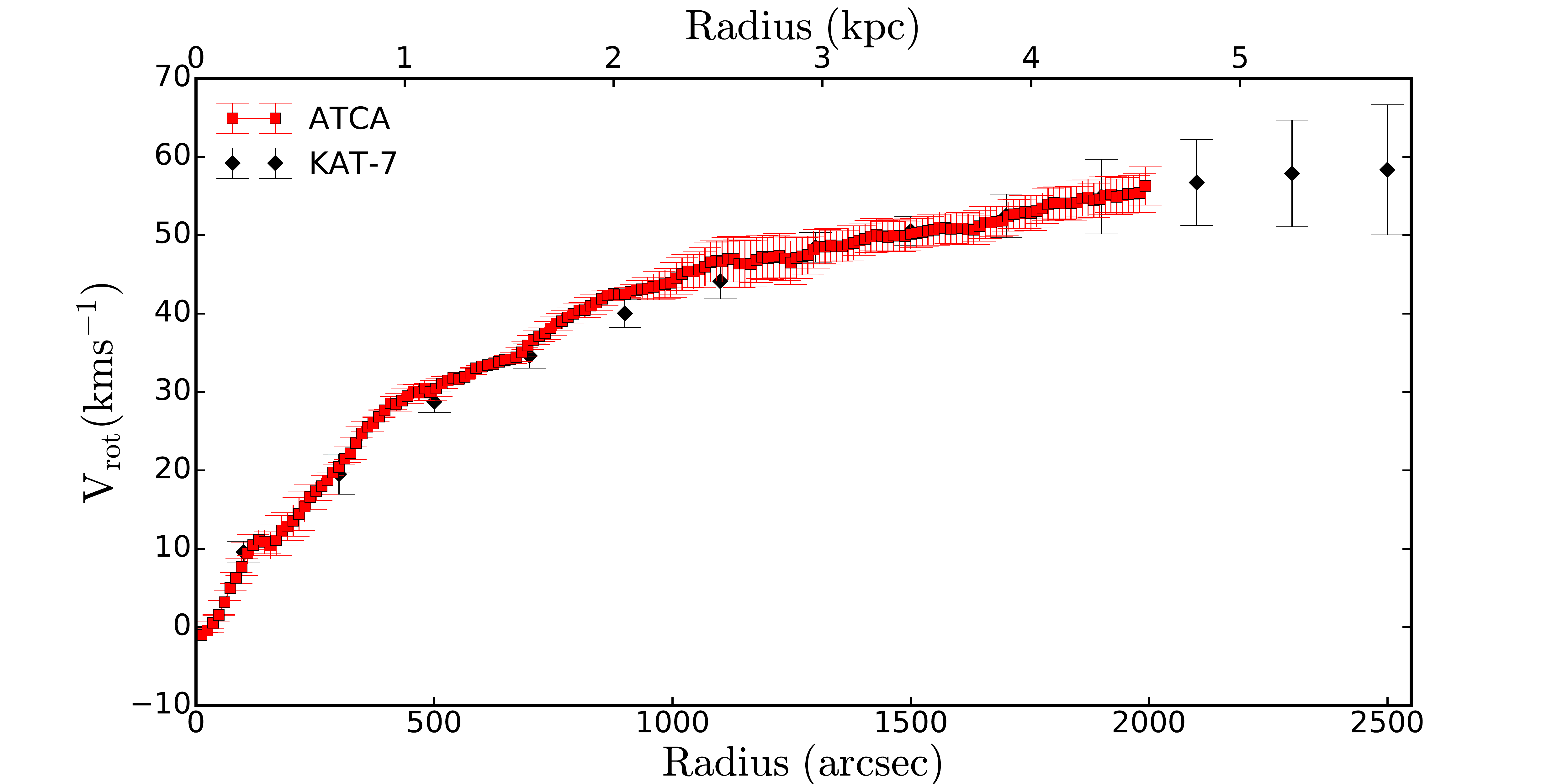}
  \caption{Comparison between the KAT-7 (black diamonds) and the ATCA (red squares) rotation curves.}
   %The inset histogram show the distribution of the difference whereby the dash vertical lines indicate $|\Delta{z}|\,/(\,1\,+\,z_{spec})\,=\,0.2$. 
 % The mean $(\mu)$, and standard deviation $(\sigma)$ of the distribution are indicated in the top right of the panel}
  \label{fig_speczvsphotz} 
\end{figure}

\begin{figure}
  \includegraphics[width = \columnwidth]{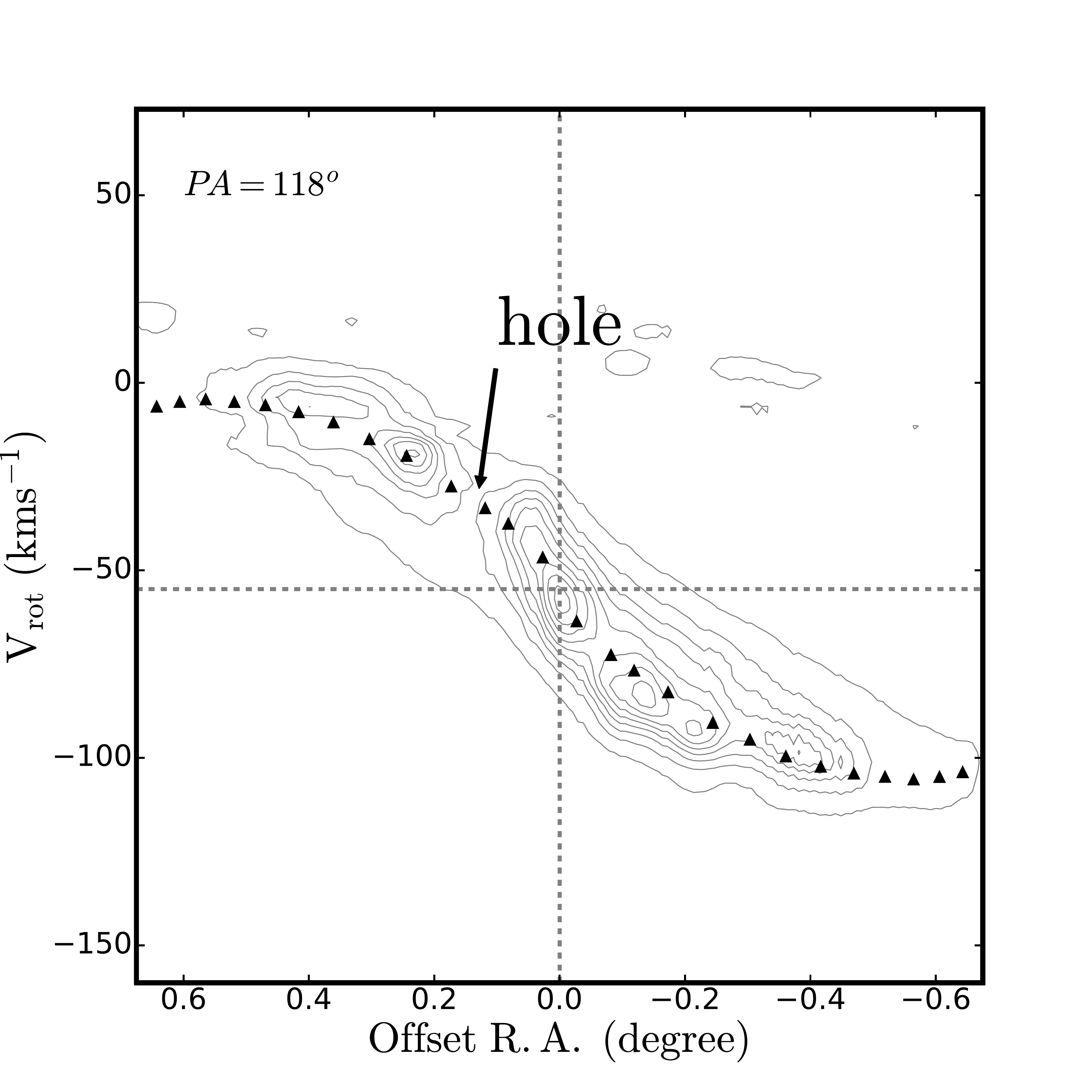}
  \caption{Position velocity diagram of NGC 6822 with the projected rotation curve over plotted. The dotted grey lines represent the centre of the galaxy and the systemic velocity. Superimposed is the rotation curve derived from the tilted ring model, corrected for the inclination of a slice along the galaxy major axis.}
   %The inset histogram show the distribution of the difference whereby the dash vertical lines indicate $|\Delta{z}|\,/(\,1\,+\,z_{spec})\,=\,0.2$. 
 % The mean $(\mu)$, and standard deviation $(\sigma)$ of the distribution are indicated in the top right of the panel}
  \label{fig_speczvsphotz} 
\end{figure}

\begin{figure*}
\centering
   \subcaptionbox{Observed velocity field\label{fig3:a}}{\includegraphics[width=2.2in]{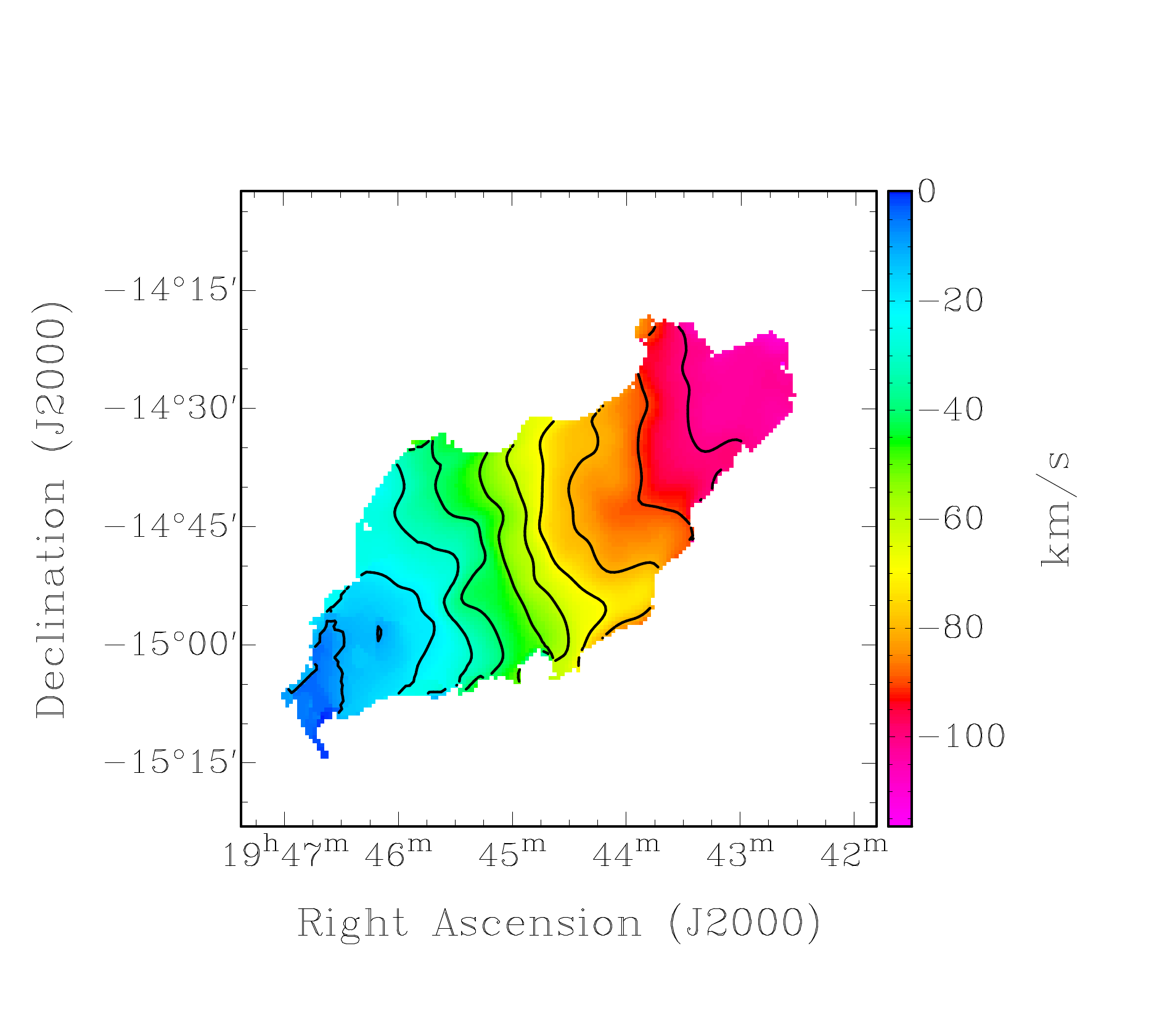}} \\ %
   \subcaptionbox{Model velocity field (both)\label{fig3:b}}{\includegraphics[width=2.2in]{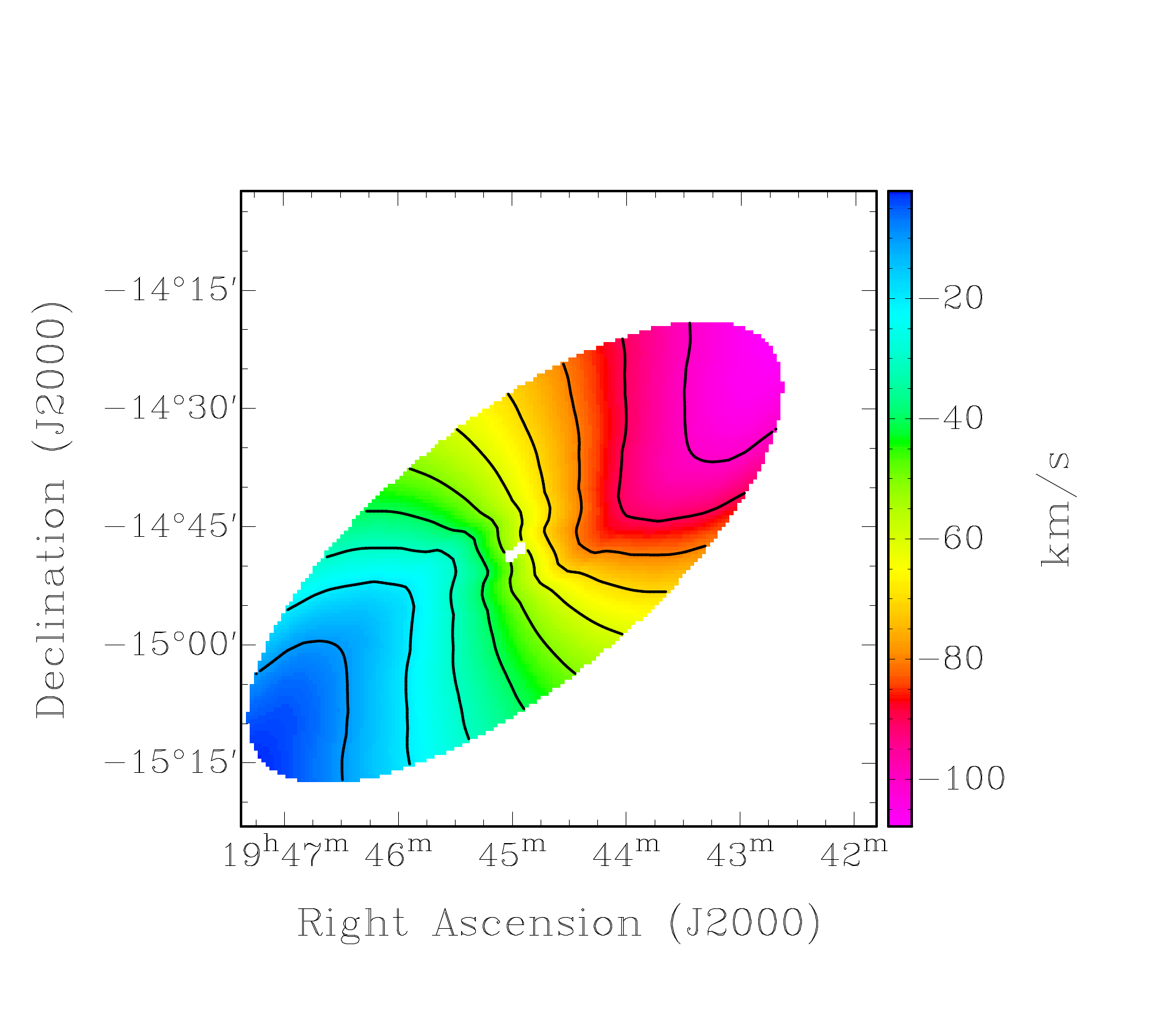}} \hspace{-3mm}  
   \subcaptionbox{Model velocity field (approaching)\label{fig3:a}}{\includegraphics[width=2.2in]{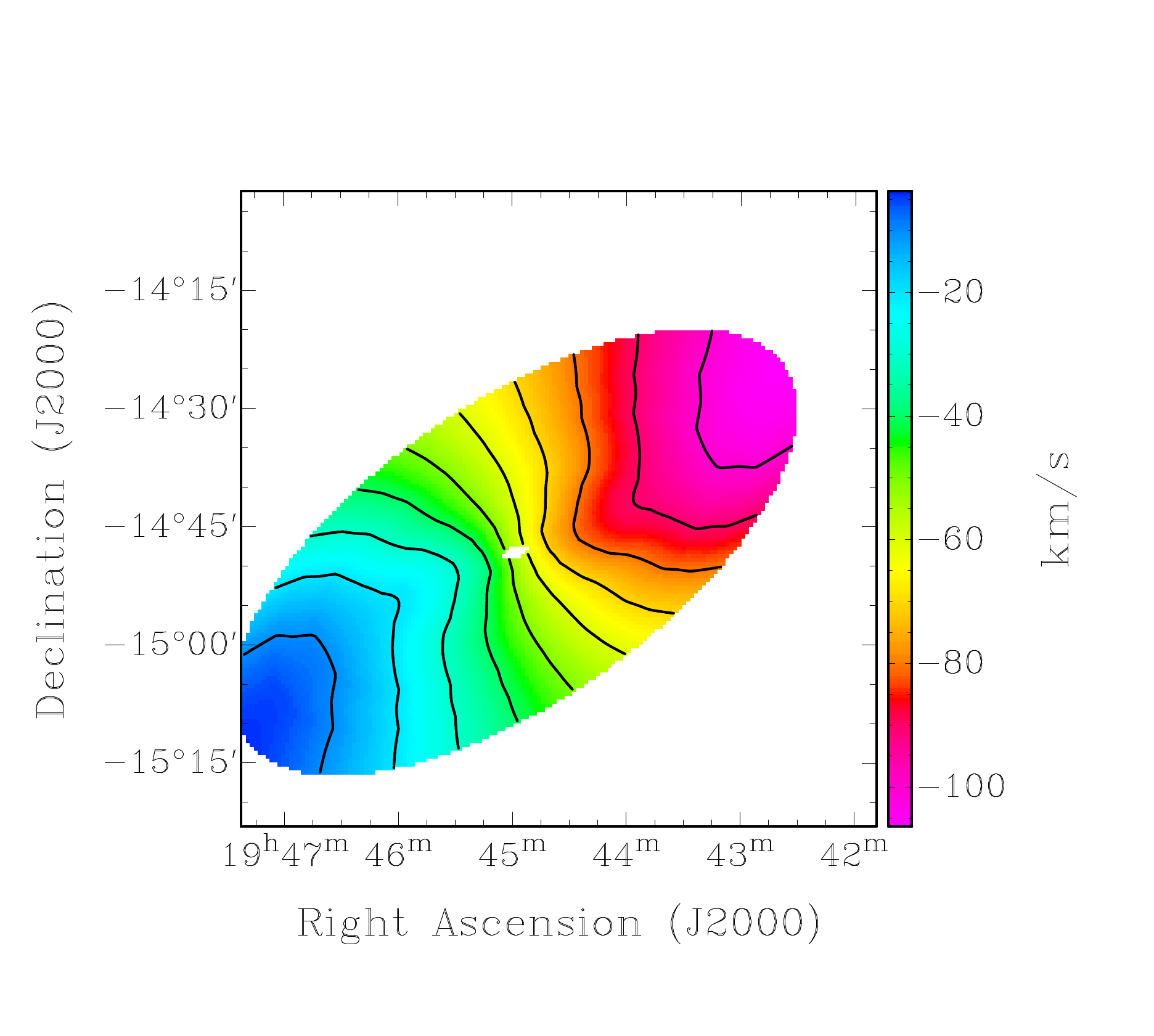}}\hspace{-3mm}%
   \subcaptionbox{Model velocity field (receding) \label{fig3:b}}{\includegraphics[width=2.2in]{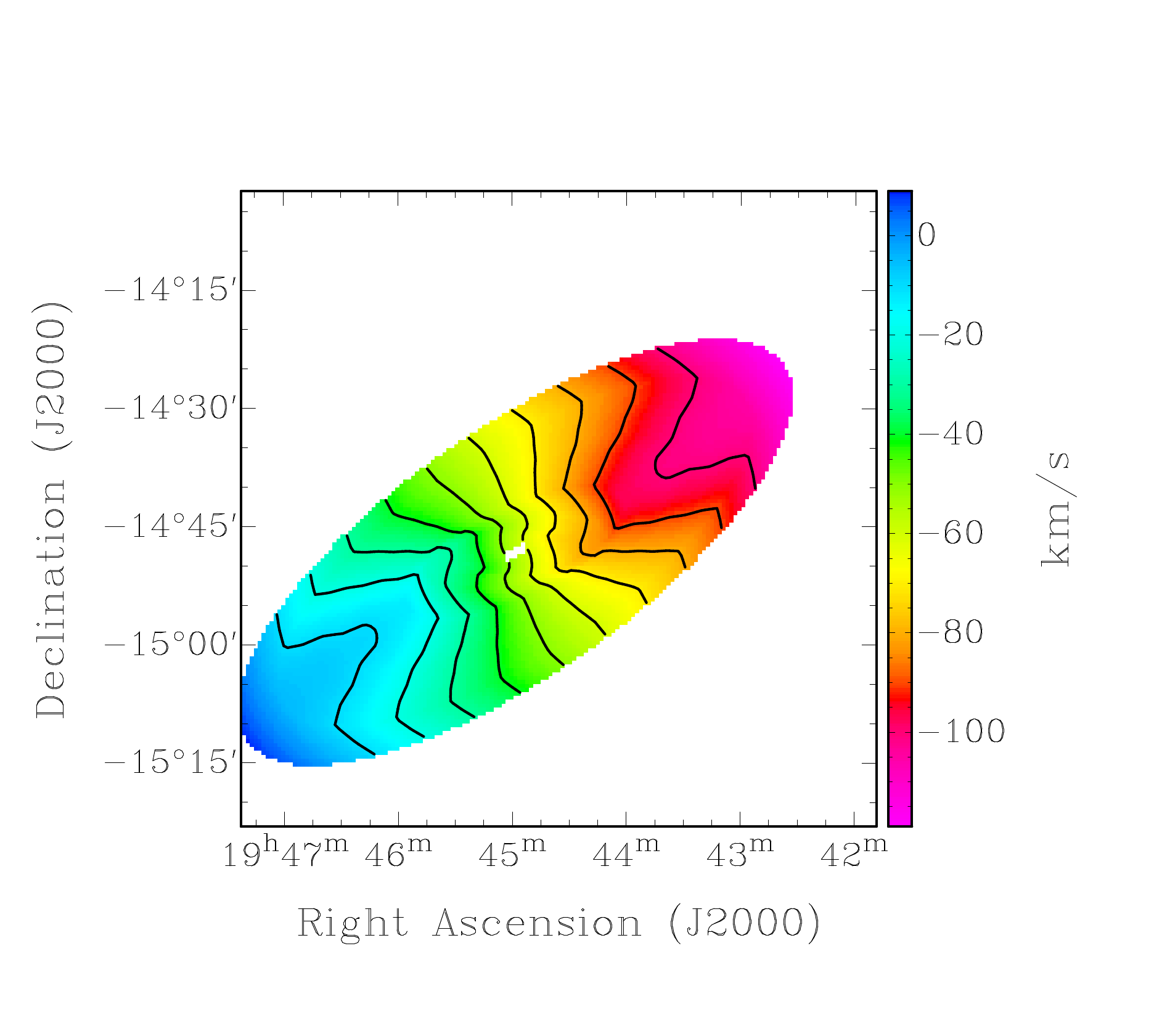}}\hspace{-3mm} \\%
   \subcaptionbox{Residual (both) \label{fig3:b}}{\includegraphics[width=2.2in]{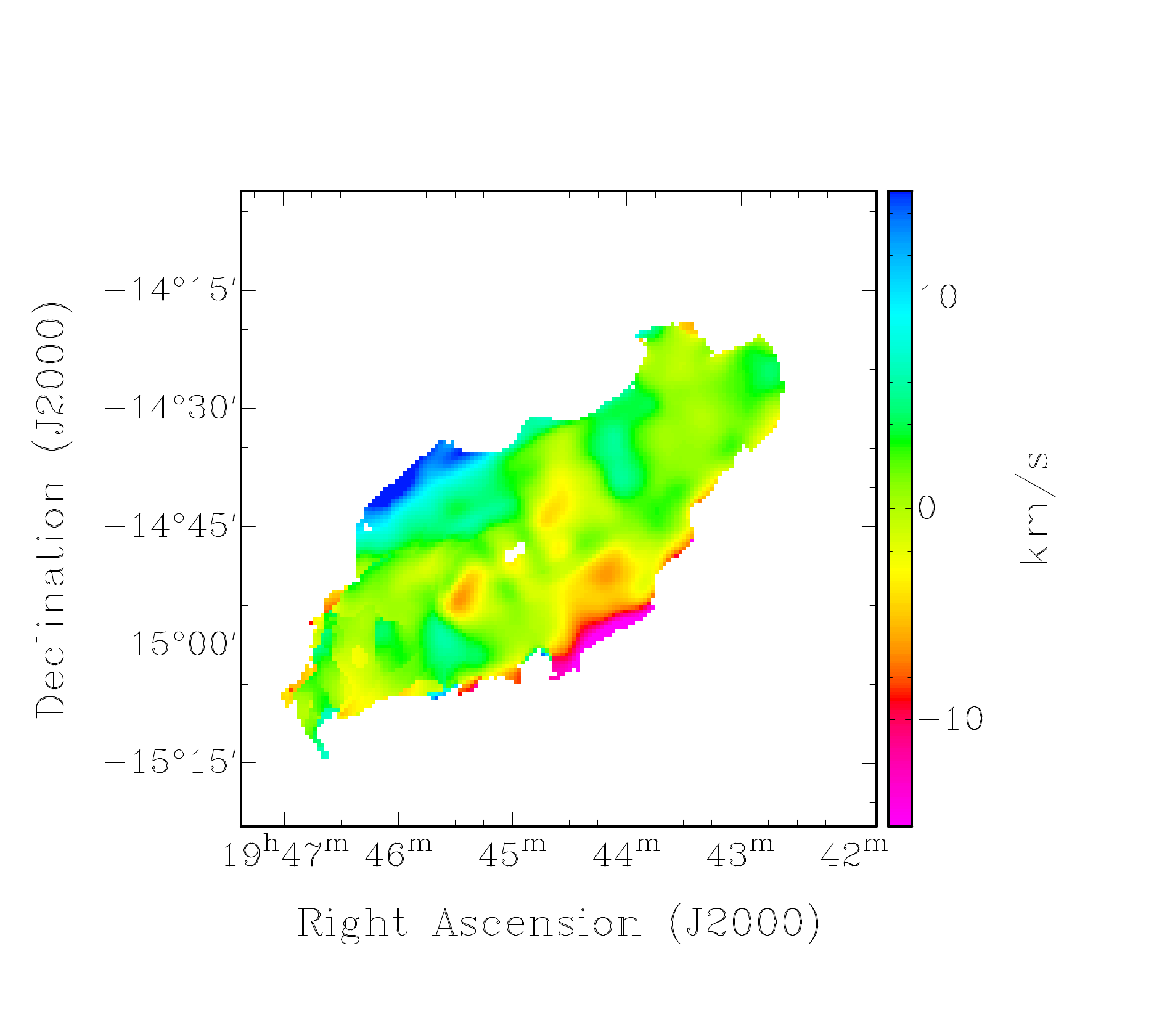}} \hspace{-3mm}  
   \subcaptionbox{Residual (approaching) \label{fig3:a}}{\includegraphics[width=2.2in]{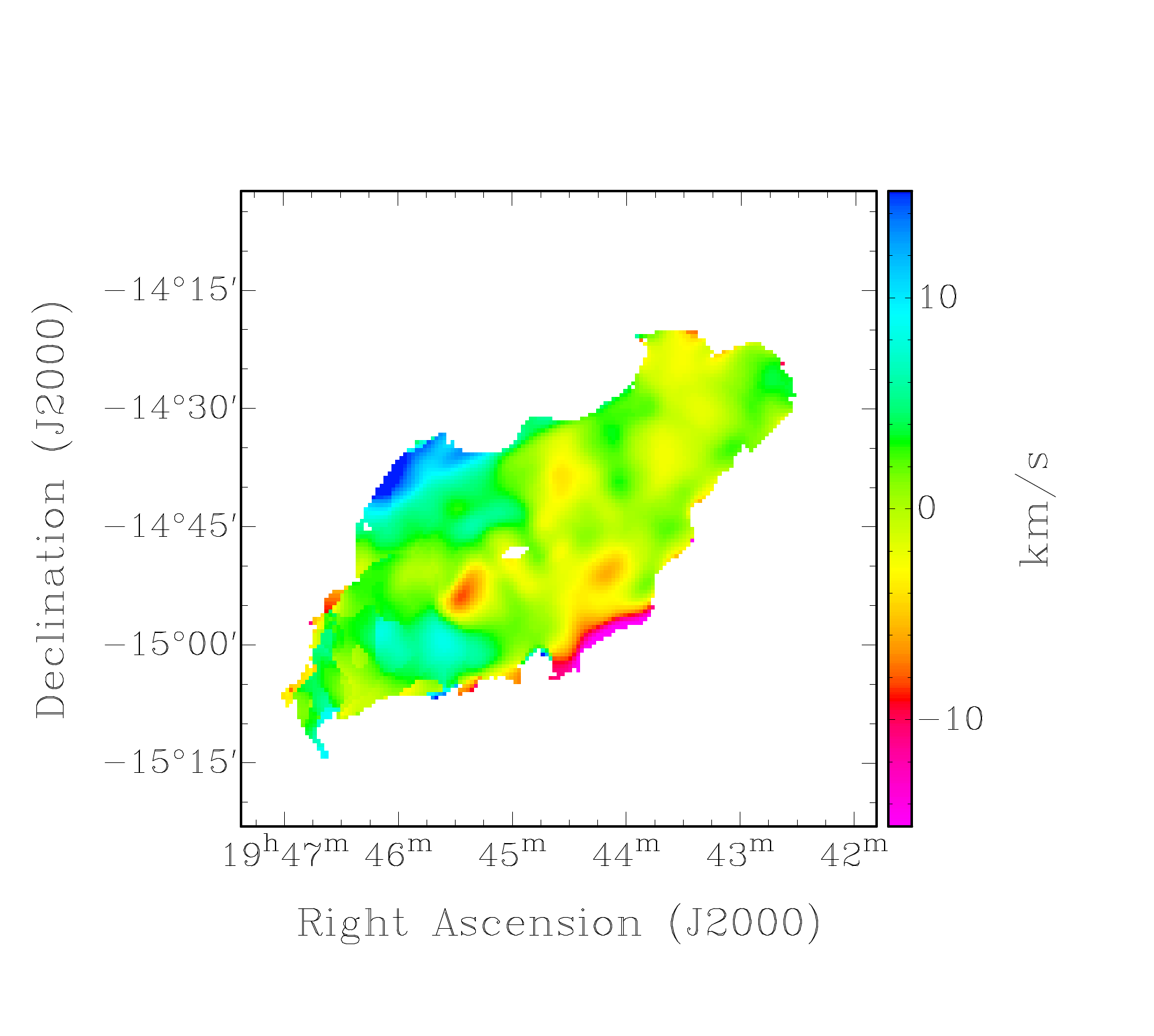}}\hspace{-3mm}%
   \subcaptionbox{Residual (receding) \label{fig3:b}}{\includegraphics[width=2.2in]{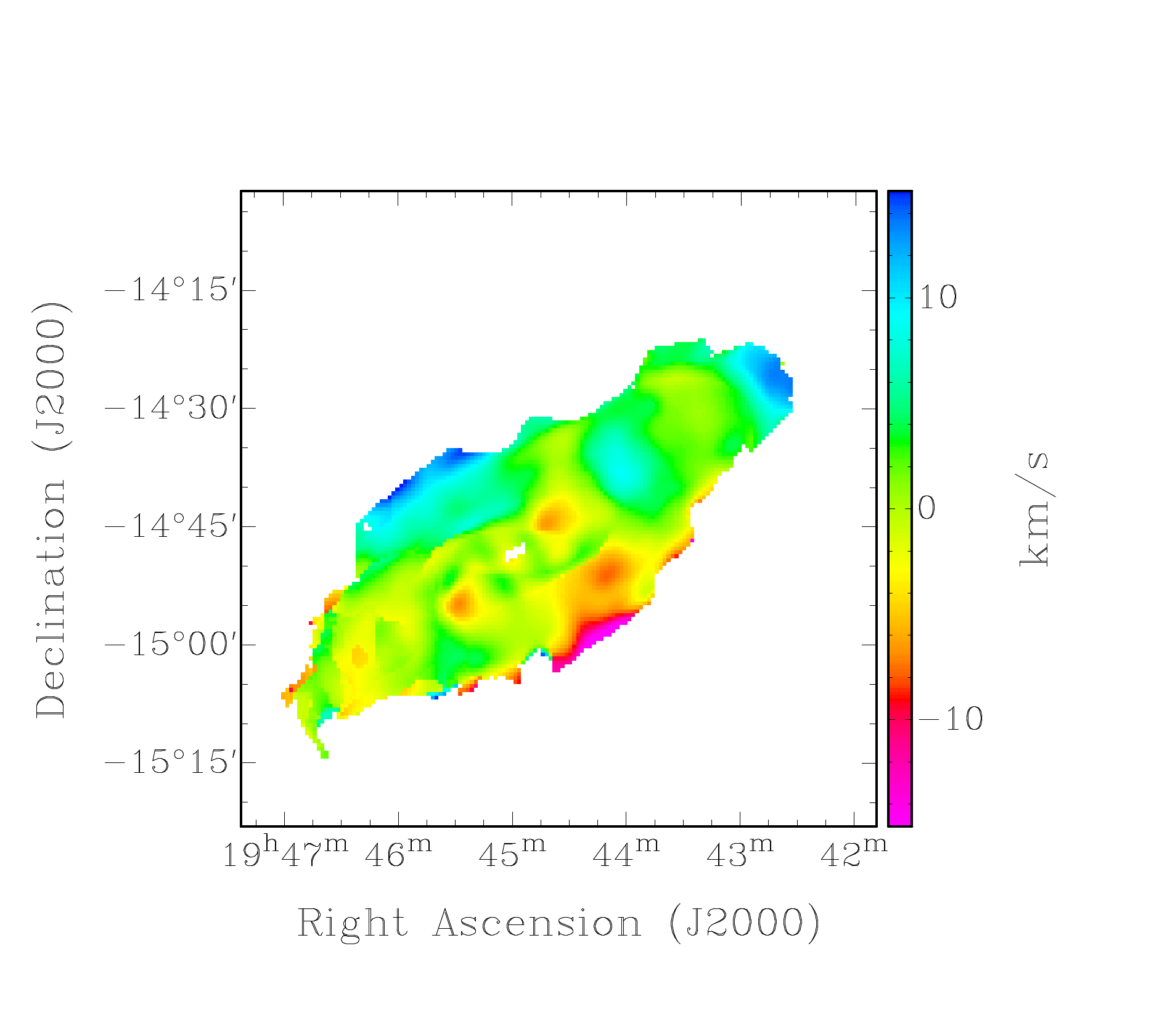}}\hspace{-3mm} \\% 
   \caption{Comparison of the observed velocity field top panel (a) and the circularly symmetric model velocity fields (middle panel) derived from the tilted ring fits for both (b), approaching (c), and receding (d) sides. The contours for the velocity fields and model velocity fields run from -100 to 10 kms$^{-1}$ in steps of 10 kms$^{-1}$. The bottom panel shows the residual velocity fields (observed-model) for both (e), approaching (f), and receding (g) sides.}  
   \end{figure*}

\begin{table}   
%\scriptsize
\caption{\small Radial Variation of the H\textsc{i} Surface Densities $\Sigma_{g}$, the gas Velocity Dispersion $\sigma$ and the Rotation Velocity for the KAT-7 Data of NGC 6822 from moment analysis.}
\begin{minipage}{\textwidth}
\begin{tabular}{l@{\hspace{0.30cm}}c@{\hspace{0.30cm}}c@{\hspace{0.30cm}}c@{\hspace{0.30cm}}c@{\hspace{0.30cm}}c@{\hspace{0.30cm}}c@{\hspace{0.30cm}}}   
\hline

Radius &$\Sigma_{g}$ &  $\sigma $& V$_{0}$&$\Delta$ V & $\sigma$/V & V$_{\text{{c}}}$  \\
arcsec &M$_{\odot}$pc$^{-2}$ & kms$^{-1}$& kms$^{-1}$&kms$^{-1}$ & $\%$ & kms$^{-1}$ \\ \hline \hline
100&2.5&10.7&7.2&1.6 &140&9.6\\
300&4.4&10.4&19.0&2.3&54&19.5\\
500&6.6&9.8&28.4&1.9&34&28.8\\
700&7.8&8.8&34.0&2.1&25&34.6\\
900&5.6&8.4&39.2&1.6&21&40.1\\
1100&2.3&7.9&42.9&2.4&18&44.1\\
1300&1.7&7.2&46.6&2.3&15&48.5\\
1500&1.3&5.6&48.9&2.1&11&50.5\\
1700&0.9&5.2&51.4&2.5&10&52.5\\
1900&0.6&4.7&53.8&2.9&9&54.9\\
2100&0.3&3.6&55.5&3.8&7&56.7\\
2300&0.2&2.7&56.3&4.6&5&57.8\\
2500&0.1&1.0&57.7&6.8&2&58.3\\

%\multicolumn{6}{@{} p{8.5 cm} @{}}{\footnotesize{\hspace{3cm} VLT/NACO DATA}}\\\\

\hline    
\multicolumn{7}{@{} p{8.5 cm} @{}}{\footnotesize{\textbf{Notes.} Column (1) gives the radius, column (2) the surface densities, column (3) the velocity dispersion, column(4) the observed rotation velocities, column (5) the errors of those velocities, column (6) the ratio between the velocity dispersion and the rotation velocity, and column (7) the corrected velocities used for the mass models.}}
\label{coords_table}
 
\end{tabular}   

\end{minipage}
\end{table}  

\section{Mass models and dark matter content}
The rotation curve reflects the dynamics of the disk due to the total mass of the galaxy, luminous and dark matter. Dwarf irregulars, like most low-mass surface density galaxies are believed to be dominated by dark matter at all radii due to the small contribution of luminous matter (stars and gas) to the total dynamics \citep{1989ApJ...347..760C}. The extended H\textsc{i} rotation curves of dwarf irregulars allow us to probe dark matter potentials to much larger radii, making them ideal objects for studying dark matter properties in galaxies. To this end, we decompose the observed rotation curve of NGC 6822 into the luminous and dark 
matter mass components and verify if dark matter indeed dominates the total dynamics of systems such as NGC 6822.
\subsection{Dark matter models}

\subsubsection{Isothermal halo model}
This model is the representation of a core like halo model \citep{1991MNRAS.249..523B}. This model describes the mass distribution well (sum of dark and baryonic matter). It has the following form: 
\begin{equation}
\rho_{\text{ISO}}(R) = \frac{\rho_{0}}{1 + \left(R/R_{C} \right)^{2}},
\end{equation}
where $\rho_{0}$ and $R_{C}$ are the central density and core radius of the halo, respectively. This gives rise to the mass distribution
with a sizable constant density core ($\rho \sim  R^{0})$ at centers of galaxies. The rotation velocity induced by the mass distribution 
is given as
\begin{equation}
V_{\text{ISO}}(R) = \sqrt{4\pi G \rho_{0}R_{C}^{2}  \left[1 - \frac{R}{R_{C}} \text{arctan} \left(\frac{R}{R_{C}} \right)\right]}
\end{equation}
\subsubsection{The Navarro, Frenk, and White DM model (NFW)}
The NFW profile, also known as the universal profile \citep{1997ApJ...490..493N} is the commonly adopted DM halo profile in the context of the $\Lambda$CDM cosmology. It was derived from the N-body simulations. The density profile is given by
\begin{equation}
\rho_{\text{NFW}}(R) = \frac{\rho_{i}}{(R/R_{S})[(1 + R/R_{S})]^{2}},
\end{equation}
where $R_{S}$ is the characteristic radius of the halo and $\rho_{i}$ is related to the density of the universe at the time of collapse of 
the DM halo. This gives a cusp feature having a power mass density distribution $\rho \sim R^{-1}$ towards the centers of the galaxies. 
The corresponding rotation velocities are given by
\begin{equation}
V_{\text{NFW}} = V_{200}\sqrt{\frac{\ln(1+cx - cx/(1 + cx)}{x[\ln(1 + c) - c/(1 + c)]}},
\end{equation}
where $x$ is defined as  $R/R_{200}$. c is the concentration parameter defined as $R_{200}/R_{S}$, $V_{200}$ is the rotation velocity at the radius $R_{200}$ 
where the density contrast with respect to the critical density of the universe exceeds 200, roughly the viral radius. 
\subsubsection{Gas and Stellar components}
The H\textsc{i} gas surface density profile was computed from the observed H\textsc{i} column density map. We use the GIPSY task 
ELLINT and apply the tilted ring parameters given in section 4. The gas surface density profile of the galaxy was scaled by 1.4 to take into account Helium and other metals. The H\textsc{i} surface density profile is presented in Figure 5. We convert the gas surface density to the corresponding gas rotation velocities by using the GIPSY task ROTMOD, assuming an infinitely thin disk.  
We use the Wide Field Infrared Survey Explorer (WISE) surface brightness profiles at 3.4 micron to account for the stellar contribution. At 3.4 microns, WISE probes the emission from the old stellar disk population and is also less affected by dust. The luminosity profile is converted into stellar mass density by adopting the method described by \citet{2008AJ....136.2761O}. First the surface brightness profile in mag/arcsec$^{2}$ is converted to luminosity density in units of L$_{\odot}$/pc$^{2}$ and then converted to mass density using the following formulae: 
\begin{equation}
 \Sigma [M_{\odot} pc^{-2}] =(M/L)_{*}^{3.4} \times 10^{-0.4(\mu_{3.4}-C^{3.4})},
\end{equation}
 where (M/L)$_{*}^{3.4}$ is the stellar mass to light ratio in the 3.4 micron, $\mu_{3.4}$ is the surface brightness in mag/arcsec$^{2}$ and C$^{3.4}$ is the constant value that is used to convert mag/arcsec$^{2}$ to $L_{\odot} pc^{-2}$. 
 C$^{3.4}$is given by $M_{\odot}^{3.4}$ + 21.56, where $M_{\odot}^{3.4}$ = 3.24 is the absolute magnitude of the sun in the 3.4 micron band. The derived profile was then converted to stellar rotation velocities using the GIPSY task ROTMOD. 
\subsection{Fitting NFW and ISO models for NGC 6822}
The GIPSY task ROTMAS was used to construct the mass models for NGC 6822. We fitted the ISO and NFW models to the rotation curve derived from the tilted ring model, taking into account the mass of the luminous matter (stars and gas). In all the fittings, the gas surface densities were fixed. Inverse squared weighting of the rotation curves with uncertainties was used during the fitting procedure. 

The mass to light ratio $\Upsilon_{*}$ is one of the largest source of uncertainty in deriving the mass model. The value of $\Upsilon_{*}$ cannot be derived from the rotation curve alone. We scale the upper and lower limits of the mass to light ratio using 2 different assumptions. 1) we fix the $\Upsilon_{*}$ of the stellar component to the predetermined value of $\Upsilon_{*}$ $\approx$ 0.2 at mid-infrared band from the literature \citep{2016AJ....152..157L}, and 2) we derive a set of fits where $\Upsilon_{*}$ is left as a free parameter. In this case we let the fitting program choose its best value. 
\subsection{Results from the ISO and NFW DM halo fits}
The fitted parameters for the mass models are summarized in Table 4 and Figure 10 for both the ISO and the NFW halo models. We find that the ISO halo model provides a better fit than the NFW model in all cases. The NFW model fails to fit the rotation curve irrespective of the assumption of $\Upsilon_{*}$ used. Both NFW models gave unphysical fitted parameters M/L= 0 (free) or too small value of c (model). On the other hand, the ISO halo model provides reasonable halo parameters. The best fit to the rotation curve: i.e we let $\Upsilon_{*}$ be a free parameter in the fit produces the smallest value of the reduced $\chi^{2}$. We tried to constrain the maximum value of $\Upsilon_{*}$ by scaling the stellar rotation curve to contribute maximally to the observed rotation curve. In this case, sensible results are obtained when the $\Upsilon_{*}$ value is lowered to the predetermined value of 0.2. This shows that NGC 6822 is not a maximum disc galaxy. Our derived M/L values of NGC 6822 are consistent with the literature. \citet{2003MNRAS.340...12W} derived the best fit M/L ratio of 0.10 $\pm$ 0.13 and 0.35 $\pm$ 0.04 for NGC 6822. The small derived M/L ratio confirms that the stellar component of NGC 6822 has very little impact on the total mass of the system. It is clear from Figure 10 that dark matter is the most massive component contributing to the total rotation curve. This makes NGC 6822 dark matter dominated even in the rising part of the rotation curve (for R > 1 kpc). 

The distribution of dark matter in NGC 6822 is typical of most dwarf irregular galaxies when compared with the literature (e.g.,\ \citealp{2008AJ....136.2648D,2011AJ....141..193O}) show that most dwarf irregular galaxies in their sample are better described by a core-like model dominated by a central constant-density core with small M/L ratios. 
%Our derived mass model results of NGC 6822 show a similar trend to the results from the literature on dwarf irregular galaxies. fits of most dwarf irregulars from the literature. \citet{2003MNRAS.340...12W} mass model fit of NGC 6822 produces good fits with the ISO halo model while 
%\begin{figure*}[htp]
%\centering
 % \subcaptionbox{3a\label{fig3:a}}{\includegraphics[width=3.5in]{isothermal.eps}}\vspace{5 mm} 
  %  \subcaptionbox{3b\label{fig3:b}}{\includegraphics[width=3.5in]{isothermal.eps}}
  %\vspace{2\baselineskip}
  %\end{figure}

\begin{figure}
\centering
\resizebox{1.0\hsize}{!}{\rotatebox{0}{\includegraphics{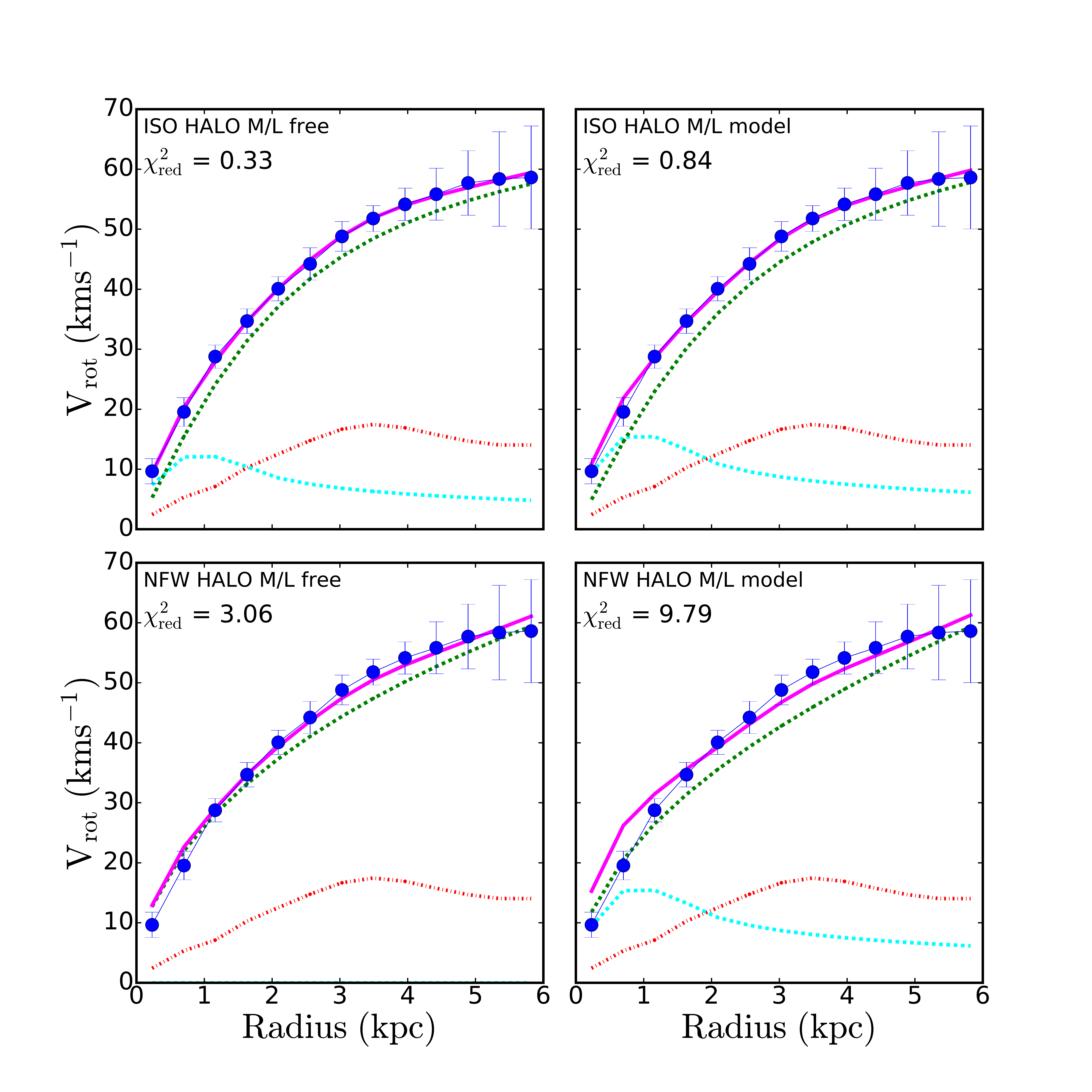}}}
\caption{\small {ISO and NFW mass modeling results of NGC 6822. The decomposition of NGC 6822 rotation curve using two assumption of $\Upsilon_{*}$. The blue circles indicate the observed rotation curve, the magenta lines show the fitted rotation curve, the green dotted lines indicate the dark matter rotation velocities, and the red dot-dashed and blue dashed lines show the rotation velocities of the gas and of the stellar components respectively}.}
\label{distrelation}
\end{figure}

\begin{table}
\scriptsize

\caption{\small Results for the Mass Models of NGC 6822.}
\begin{minipage}{\textwidth}
\begin{tabular}{l@{\hspace{0.6cm}}c@{\hspace{0.6cm}}c@{\hspace{0.5cm}}c@{\hspace{0.5cm}}c@{\hspace{0.5cm}}}   
\hline

%Model &Parameter & Result &sa &fr\\
                         %&(sec)       &(Mpc) & (mag) & ($L_{\odot}$) &($\rm M_{\odot}yr^{-1}$)  \\
                %~~~~~~(1)    &   (2)        &  (3)   \\         
   
\hline \hline  
%\multicolumn{6}{@{} p{8.5 cm} @{}}{\footnotesize{\hspace{3cm} VLT/NACO DATA}}\\\\
&&ISO Halo&& \\
\cline{2-3}
Assumption &$\Upsilon_{*}$ & R$_{c}$ &$\rho_{0}$ &$\chi_{\text{red}}^{2}$\\ 
(1)&(2)&(3)&(4)&(5) \\ \hline
M/L free &0.12& 1.87 $\pm$ 0.07& 29.22 $\pm$1.38&0.33 \\
M/L model & 0.20& 2.06 $\pm$ 0.09& 25.59$\pm$ 1.40&0.84\\ \hline
&&NFW Halo&& \\
\cline{2-3}
Assumption &$\Upsilon_{*}$ & c &R$_{200}$ &$\chi_{\text{red}}^{2}$\\ 
(6)&(7)&(8)&(9)&(10) \\ \hline
M/L free &0.00& 3.34 $\pm$ 1.14& 165.39 $\pm$54.39&3.06 \\
M/L model & 0.20& 0.01$\pm$ ....& 1120.71$\pm$ ....&9.79\\ \hline

\multicolumn{5}{@{} p{9.0 cm} @{}}{\footnotesize{\textbf{Notes.} Columns 1 and 6, the stellar $\Upsilon_{*}$ assumption. Column 2 and 7, 
$\Upsilon_{*}$ . Column 3, fitted core radius of the ISO halo model(10$^{-3}$ M$_{\odot}$pc$^{-3}$). Column 4, fitted core density of the pseudo-isothermal halo model (10$^{-3}$ M$_{\odot}$ pc$^{-3}$). Column 9, the radius in kpc where the density contrast exceeds 200. Column 8, concentration parameter c of the NFW halo model. Columns 5 and 10, reduced $\chi^{2}$ value. The dotted line (...) are due to unphysical large values of uncertainties.}}
\label{coords_table}
 
\end{tabular}   

\end{minipage}
\end{table}  
\subsection{MOND Models for NGC 6822}
A alternative to dark matter is the modified Newtonian dynamics (MOND)\citep{1983ApJ...270..384M,1988ApJ...333..689M}. MOND is a phenomenological modification of Newton's law of gravitation which produces the dynamics of galaxies, without the need for additional dark matter. Only the stellar and gas contributions are needed to explain the rotation curve.
\subsubsection{MOND Models Using the $Standard$ Interpolation Function}
The standard interpolating function \citep{1983ApJ...270..384M} is given as
\begin{equation}
\nu(x) = \frac{x}{\sqrt{(1 + x^{2})}}.
\end{equation}
For x $\ll$ 1 the system is in deep MOND regime with $g = (g_{N}a_{0})^{1/2}$ and for x $\gg$ 1 the gravity is Newtonian. The MOND rotation curves is given by
\begin{equation}
V_{\text{rot}}^{2} = \frac{V^{2}_{\text{sum}}}{\sqrt{2}} \sqrt{ 1 + \sqrt{1 + 2(2ra_{0}/V^{2}_{\text{sum}})^2}}.
\end{equation}
where 
\begin{equation}
V^{2}_{\text{sum}} = V_{b}^{2} + V^{2}_{d} + V^{2}_{g},
\end{equation}
where $V_{b}$, $V_{d}$, and $V_{g}$ are the contributions from the bulge, the disk, and the gas to the rotation curve. In the case of the Irregular galaxy 
NGC 6822, there is no bulge to consider. 
\subsubsection{MOND Models Using the $Simple$ Interpolation Function}
The simple interpolation function \citep{2005MNRAS.363..603F} is given by
\begin{equation}
\nu_{x} = \frac{x}{1 + x}.
\end{equation}
Using the same procedure as in the previous Section, we obtain the rotation velocities:
\begin{equation}
V_{\text{rot}}^{2} = \sqrt{V_{b}^{2} + V_{d}^{2} + V_{g}^{2}} * \sqrt{a_{0} *r + V_{b}^{2} + V_{d}^{2} + V_{g}^{2}}.
 \end{equation}
\citet{2005MNRAS.363..603F} found that the simple interpolating function gives more plausible M/L{\small s} compared to the standard function. 
\subsection{MOND fits and results of NGC 6822}
The MOND fitting procedure has one free parameter. We fit the rotation curve with a$_{0}$ fixed to the universal constant of 1.21 $\times 10^{-8}$ cms$^{-2}$ while letting M/L free for the standard and the simple interpolation functions. Results for  MOND fits are summarized in Table 5 and Figure 11. We see that with the using the universal constant a$_{0}$ the MOND fit fails to reproduce the distribution of matter in NGC 6822. 

\begin{figure}
\centering
\resizebox{1.0\hsize}{!}{\rotatebox{0}{\includegraphics{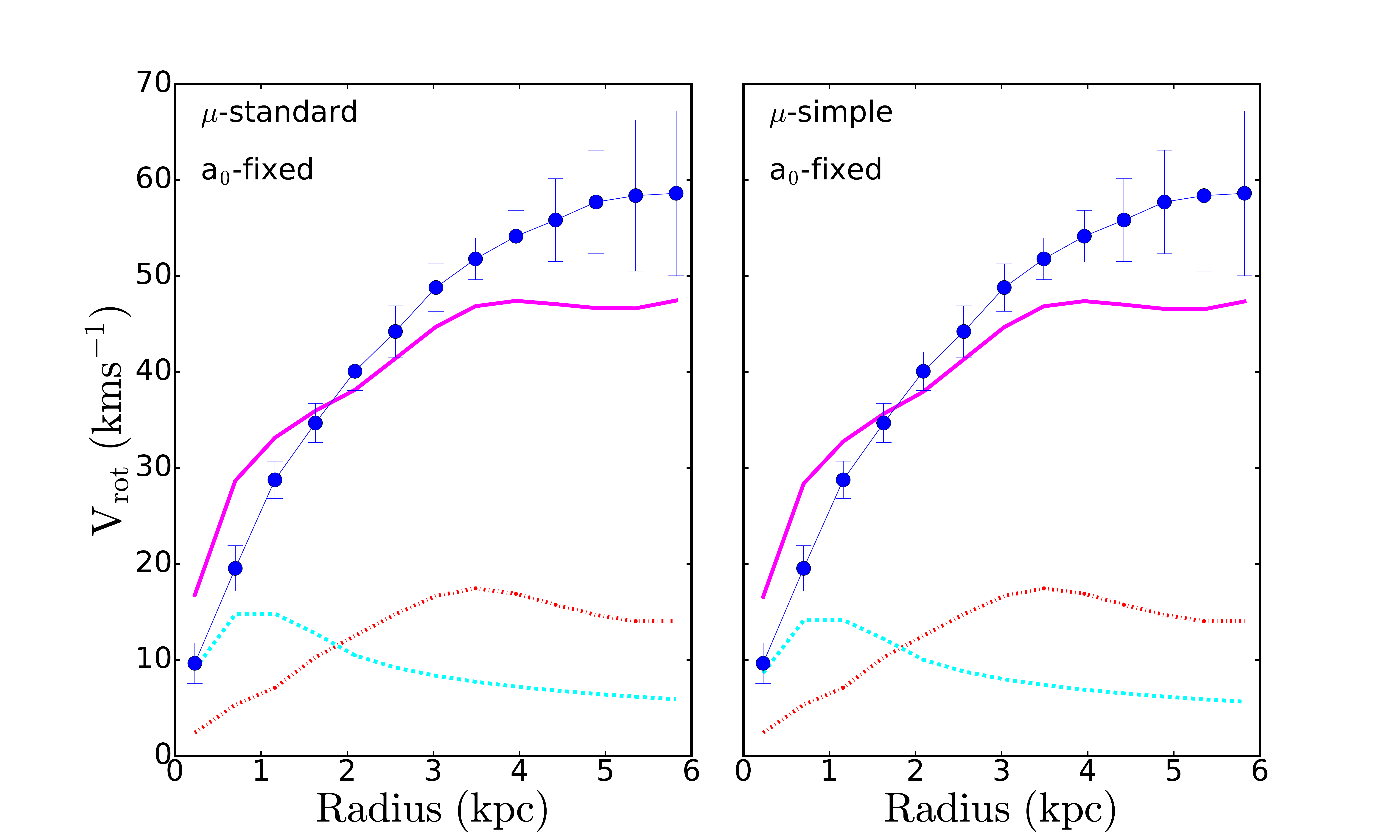}}}
\caption{MOND mass models with a$_{0}$ fixed for the standard (left) and the simple (right) interpolation functions for NGC 6822 using the derived rotation curve. The red dot-dashed curve is for the H\textsc{i} disk, the dashed light blue curve is for the stellar disk, and the continuous purple curve is the MOND contribution.}
\label{distrelation}
\end{figure}

\begin{table}   
%\scriptsize

\caption{\small Results for the MOND Models of NGC 6822 for the KAT-7 Data. The value of a$_{0}$ is fixed as 1.21 $\times$ 10$^{-8}$ cms$^{-2}$ for both models.}
\begin{minipage}{\textwidth}
\begin{tabular}{l@{\hspace{0.90cm}}c@{\hspace{0.90cm}}c@{\hspace{0.90cm}}c@{\hspace{0.90cm}}}   
\hline 
$\mu$ &  Parameter& Result \\
& & &\\ \hline \hline
Standard&(M/L)&0.18 $\pm$ 0.10\\ 
& $\chi^{2}_{\text{red}}$& 59.50\\ 
\multicolumn{2}{c}{}&\\
\cline{1-4}
Simple& (M/L)& 0.17 $\pm$ 0.09\\
&$\chi^{2}_{\text{red}}$&59.29\\
\multicolumn{2}{c}{}&\\ \hline

%\multicolumn{5}{@{} p{8.5 cm} @{}}{\footnotesize{\textbf{Notes.} Col\,(1): {\tt IRAS} survey name \textcolor{red}{with the old NACO data marked with an asterisk}; (2): Total exposure time; (3): Luminosity distance from NED Database; (4):  $K_S$-band absolute magnitude of the brightest cluster; (5): Galaxy IR luminosity from \citet{2003AJ....126.1607S}, any value marked by $\dagger$ is estimated by using the method described in $\S$\,\ref{sec_relation}; (6): SFR derived from Eq.\,\ref{Kennicut}}}
\label{coords_table}
 \end{tabular}   
\end{minipage}
\end{table}  
\section{Star formation threshold}
A threshold in the gas surface density has been inferred from the relationship between the optical edges of galaxies and their H\textsc{i} column densities. The concept of the star formation threshold is usually explained in terms of the Toomre-$Q$ parameter \citep{1989ApJ...344..685K}. The parameter is defined as
\begin{equation}
Q(r) = \frac{c_{s} k}{\pi G \Sigma_{g}},
\end{equation}
where $c_{s}$ is the sound speed in the gas, $G$ is the gravitational constant, $\Sigma_{g}$ is the gas surface density and $k$ is the epicyclic frequency, defined as:
\begin{equation}
k^2 = 2 \Bigg( \frac{V^2}{R^2} + \frac{V}{R} \frac{dV}{dR} \Bigg) s^{-2}, 
\end{equation}
where $V$ is the rotation velocity in kms$^{-1}$, and $R$ is the radius in parsec. \citet{1989ApJ...344..685K} found that a modified Toomre (1984) Q criterion, a simple thin disk gravitational stability condition, could satisfactorily describe the star formation threshold gas surface densities in active star forming galaxies. For a thin isothermal gas disk, instability is expected if the surface density exceeds a critical value: 
\begin{equation}
\Sigma_{c}= \alpha \frac{\sigma k}{ \pi G},
\end{equation}
where $\sigma$ is the gas velocity dispersion and $\alpha$ is a dimensionless constant close to unity which is included to account for a more realistic disk. The $\alpha$ constant is fitted to the threshold values of the radially varying quantity
\begin{equation}
\alpha = \Sigma_{g}/\Sigma_{c}.
\end{equation}
\citet{1989ApJ...344..685K} adopted a constant velocity dispersion of 6 kms$^{-1}$ and derived the $\alpha$ value of 0.63. If $\Sigma_{g}$ exceeds $\Sigma_{c}$ then the disk will be unstable to axisymmetric disturbances and large scale star formation can occur. The minimum density at which gas becomes unstable is a function of $k$ and $\sigma$ of the gas, and so the threshold density for cloud formation, which leads to star formation, can vary with radius in the galaxy.

\citet{1998ApJ...493..595H} used the Toomre criterion and measured a mean value of $\alpha$ $\approx$ 0.3 in irregular galaxies, a factor of two lower than that found by \citet{1989ApJ...344..685K} for spiral galaxies. This implies that the gas in irregular galaxies is more stable than the gas in spiral galaxies. Although part of this difference is due to the different velocity dispersion used, they explain that for dwarf irregulars with close to solid body rotation curves, the local shear rate, rather than the Coriolis force (essentially $k$) best describes the destruction rate of giant clouds. This implies that for irregular galaxies such as NGC 6822, the threshold stability criterion would have to be modified to take into account the local shear rate. This threshold based on the local shear rate \citep{1998ApJ...493..595H}, is described by the Oort's constant A, 
\begin{equation}
A = -\frac{1}{2}R\frac{d \Omega}{dR} = \frac{1}{2} \Bigg(\frac{V}{R} - \frac{dV}{dR} \Bigg),
\end{equation}
where d$\Omega$ is the angular speed. Then the shear critical threshold has the form 
\begin{equation}
\Sigma_{c,A} = \frac{\alpha_{A} A\sigma}{\pi G},
\end{equation}
and $\alpha_{A} \sim 2.5$

We examine the star formation threshold throughout NGC 6822 as a function of radius to determine if a subcritical gas density is preventing the galaxy from large scale star formation. We use the velocity dispersion of $\sigma$ = 7 kms$^{-1}$ which represents the median value derived from the azimuthally averaged H\textsc{i} velocity dispersion radial profile. We have assumed $\alpha$ = 1. The H\textsc{i} surface density, $\Sigma_{g}$ is multiplied by a factor of 1.4 to account for helium and other metals. We have used the H$\alpha$ images to determine the distribution of star formation in this galaxy and looked for the outer limits of star formation. H$\alpha$ emission is the primary star formation tracer, because the H$\alpha$ flux is directly proportional to the star formation rate. We use a clean calibrated H$\alpha$ map from \citet{2006AJ....131..343D} and derive the H$\alpha$ surface density using our tilted ring parameters described in Section 4. Figure 12 shows contours of H\textsc{i} superposed on the H$\alpha$ map. We see that the H$\alpha$ is found within an observed H\textsc{i} column density $\geqslant$ 10$^{21}$ cm$^{-2}$. 

From these data we have estimated the critical densities using the Toomre criteria and the cloud growth criteria based on shear which can be seen in Figure 13. At all radii, the critical densities, $\Sigma_{c}$ fails to predict star formation while the shear critical densities, $\Sigma_{c,A}$ predicts the star formation fairly well in the inner regions of the galaxy $\lesssim$3 kpc. Figure 14 shows the ratio of the observed gas surface density to the critical surface densities $\Sigma_{c}$ and $\Sigma_{c,A}$ ($\alpha$) in NGC 6822. Also plotted for comparison are the H$\alpha$ surface densities which indicate star formation rate. The star formation traced by the H$\alpha$ is seen to be higher in regions where the $\alpha$ values are high showing a similar radial trend between the star formation and $\alpha$. 

\citet{1989ApJ...344..685K} determined an average value for the parameter $\alpha$ by finding the mean value of $\Sigma_{g}/\Sigma_{c}$ at the radial limit of the observed H\textsc{ii} regions. For NGC 6822, the limit of H\textsc{ii} regions, as seen from the H$\alpha$ profile is located at a radius of $\sim$ 2.8 kpc. The mean value of $\alpha$ at that radius is $\sim$ 0.25. This is a factor 2.5 less than the value derived by \citet{1989ApJ...344..685K}. However, this value is consistent with the derived $\alpha$ values of most irregular galaxies. \citet{1998ApJ...493..595H} derived a mean $\alpha$ value of 0.26 for Sextans A, 0.30 for DDO 155, and 0.36 for DDO 168. Similar values were derived for other dwarf irregulars in their study sample. This result shows that for the Toomre-Q criterion, star formation in NGC 6822 is consistent with most dwarf irregulars.

The $\alpha$ values derived from the shear critical density show quite a different picture in the inner regions of the galaxy. The $\alpha$ values range from 0.6 to 1 within the radial limit of the H\textsc{ii} regions. The peak value around 0.7 kpc is not considered physical due to the bump in the position angle used to derive the rotation velocities (see Figure 6) and gas surface density. We derive a mean $\alpha$ value of 0.8, which is close to 1. This shows that the cloud growth criterion based on shear is better suited than the Toomre-Q criterion to explain the star formation observed in NGC 6822. Similar results have been found by \citet{1998ApJ...493..595H} for a sample of dwarf irregulars.

\begin{figure}
  \includegraphics[width = \columnwidth]{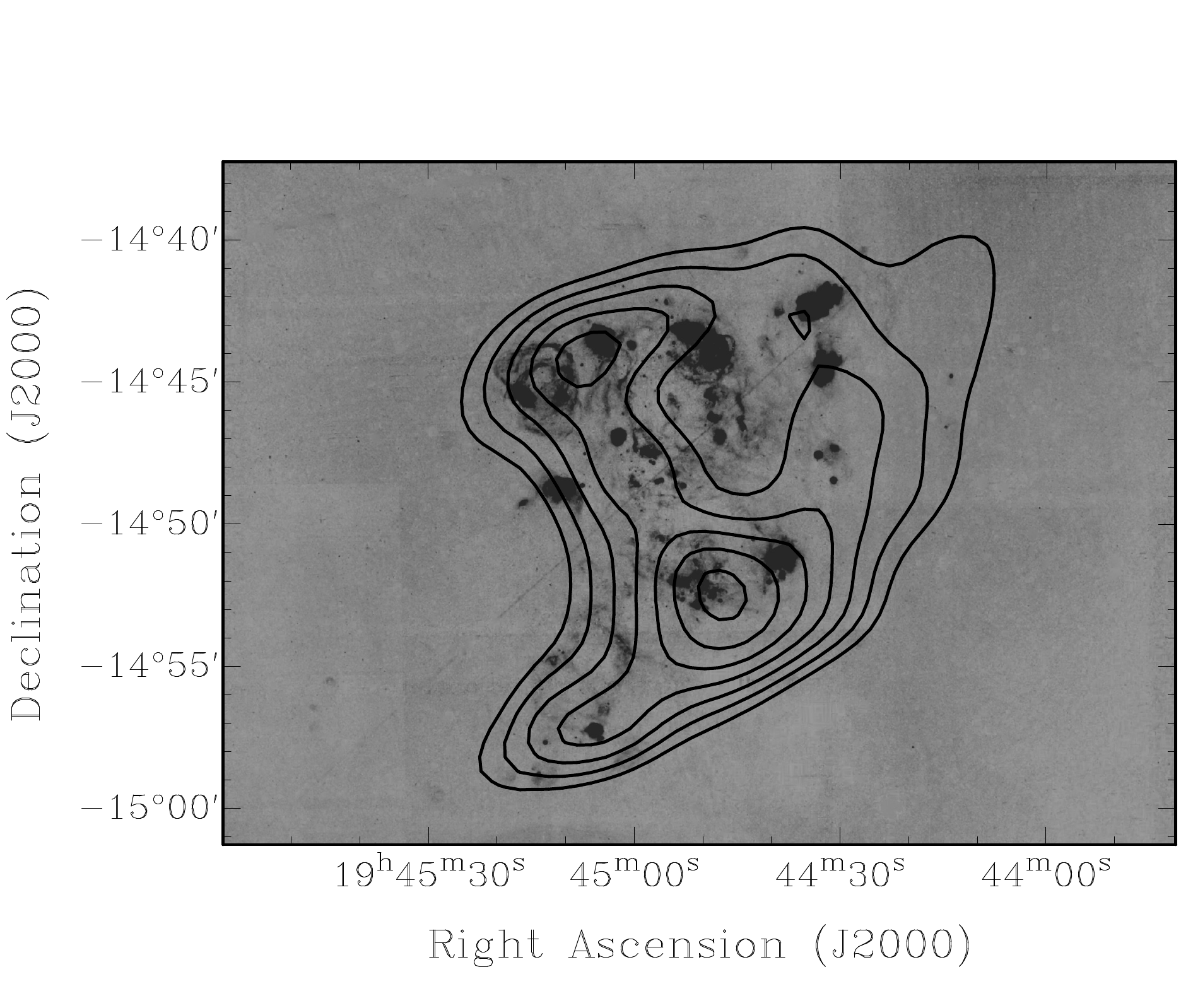}
  \caption{H\textsc{i} column density contours superposed on our H$\alpha$ image. The column density contours are 1.0, 1.2, 1.4, 1.6, 1.8, 2.0, 2.2, and 2.4 $\times 10^{21}$ cm$^{-2}$. The lowest contour covering the H$\alpha$ region is 1.0$ \times 10^{-21}$ cm$^{-2}$.}
   %The inset histogram show the distribution of the difference whereby the dash vertical lines indicate $|\Delta{z}|\,/(\,1\,+\,z_{spec})\,=\,0.2$. 
 % The mean $(\mu)$, and standard deviation $(\sigma)$ of the distribution are indicated in the top right of the panel}
  \label{fig_speczvsphotz} 
\end{figure}

\begin{figure}
\centering
\resizebox{1.0\hsize}{!}{\rotatebox{0}{\includegraphics{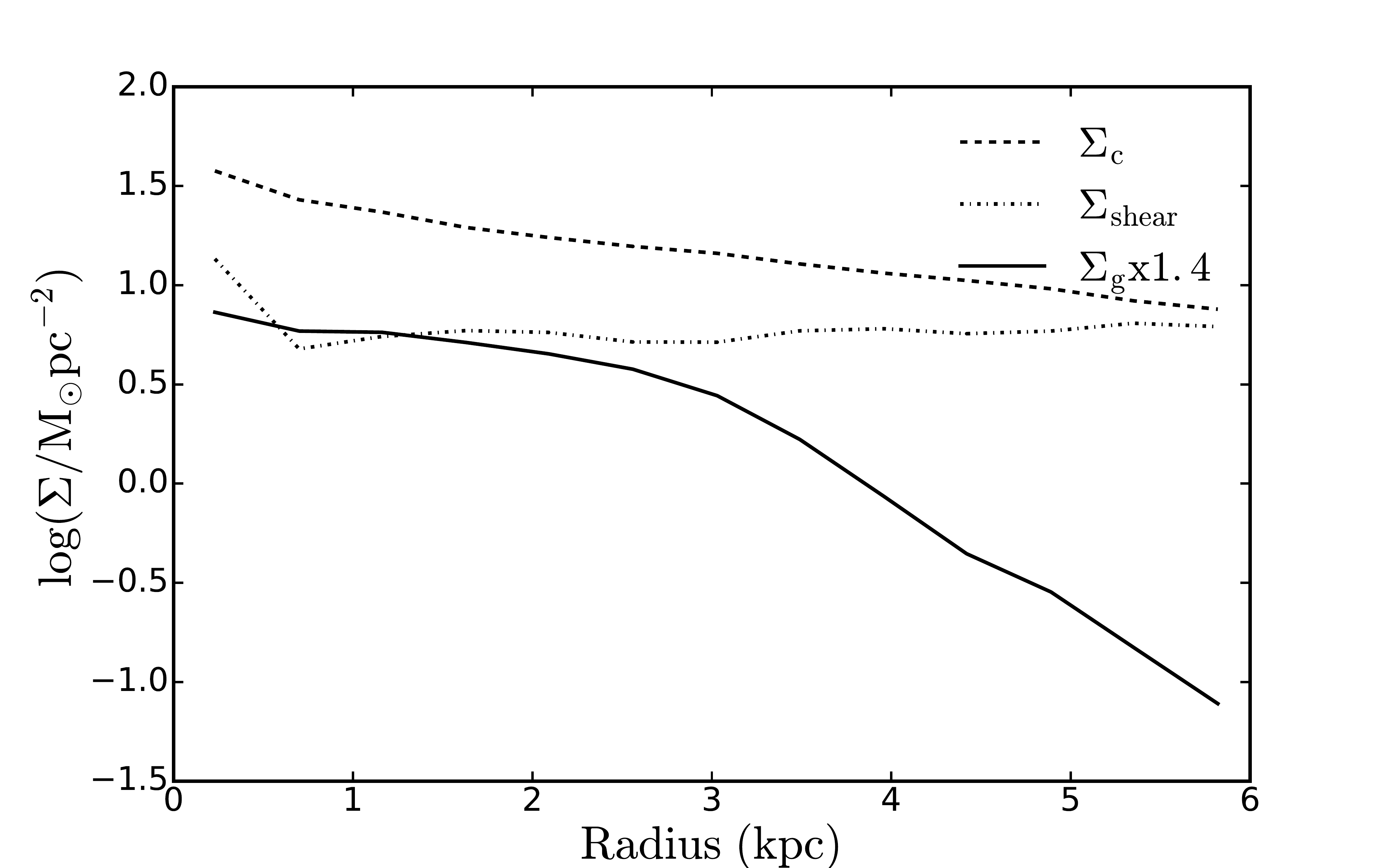}}}
\caption{\small{Comparsion of the radial distribution of the star formation threshold surface densities to the H\textsc{i} surface density. The full black line shows the gas surface density multiplied by a factor of 1.4 to correct for helium. The dashed line shows the critical density derived using the \citet{1989ApJ...344..685K} version. The dash-dotted line shows the shear critical density derived using the \citet{1998ApJ...493..595H} version}.}
\label{distrelation}
\end{figure}

\begin{figure}
\centering
\resizebox{1.0\hsize}{!}{\rotatebox{0}{\includegraphics{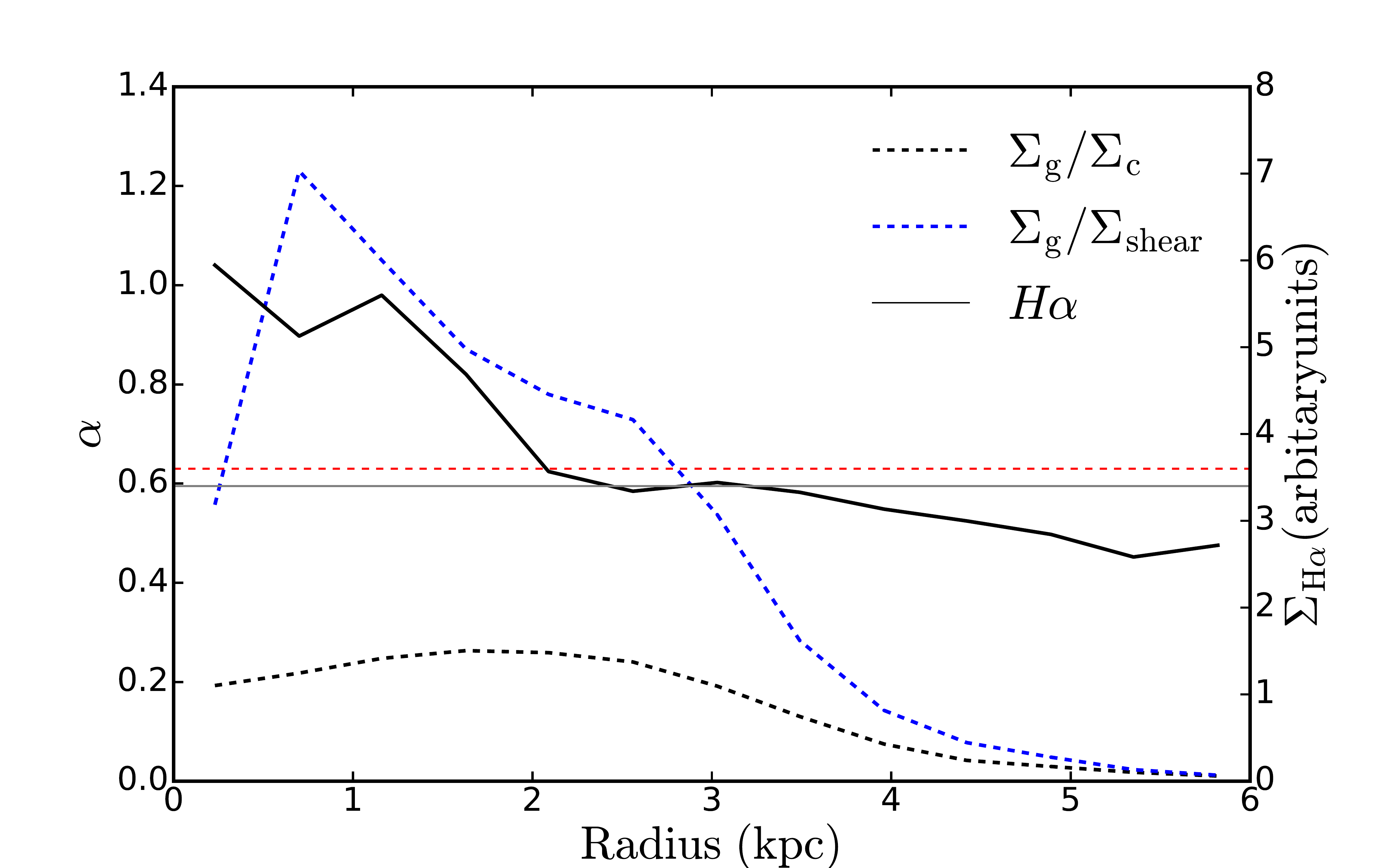}}}
\caption{\small{Radial variation in the ratio of the observed gas surface density to the critical densities derived from the Toomre-Q and cloud-growth criterion. The dashed black line shows $\Sigma_{g}/\Sigma_{c}$ while the dashed blue line shows $\Sigma_{g}/\Sigma_{c,A}$. Plotted for comparison in a black solid line are the H$\alpha$ surface densities. The dashed red line shows the median value $\alpha$ = 0.63 from \citet{1989ApJ...344..685K} above which the gas density is high enough for large scale star formation. The grey solid line represents the H$\alpha$ background noise}.}
\label{distrelation}
\end{figure}

\section{Summary}
This study presents observations of the Local Group dwarf galaxy NGC 6822, performed with the KAT-7 radio interferometer in South Africa. The KAT-7 configuration ($\sim 3.5^{\prime}$ spatial resolution and  T$_{\text{sys}} \sim $ 26 K) and the total observation time of $\sim$ 105 hrs on NGC 6822 allow us to detect the extended gas, reaching low column densities of 1$\times 10^{19}$ atoms cm$^{-2}$, which is an order of magnitude lower than the measured value for the ATCA observations. With the velocity resolution of 2.56 kms$^{-1}$ for our observations, we find that:

1. A total H\textsc{i} mass of 1.3 $\times$ 10$^{8} M_{\odot}$ is measured for NGC 6822 using the adopted distance of 0.48 Mpc. This H\textsc{i} mass, which is 23$\%$ larger than the value calculated using the ATCA observations, is surely a better estimate of the
total H\textsc{i} mass of NGC 6822 as detected by the single dish Parkes observations. We have derived the mid-point velocity of 55 $\pm$ 2 kms$^{-1}$, which corresponds to the systemic velocity found by ROTCUR. 

2. The rotation curve of NGC 6822 was derived from the velocity field map. We derive the V$_{\text{sys}}$ = 55 kms$^{-1}$, mean P.A. = 118$^{\circ}$, and mean $i$ = 66$^{\circ}$, with very little difference between the approaching and receding sides. The KAT-7 RCs agree very well with the ATCA data. Although our derived kinematical parameters are in agreement with the values derived from the ATCA observations (V$_{\text{sys}}$ = 54.4 kms$^{-1}$, mean P.A. $\sim$ 118$^{\circ}$, and  mean $i \sim 63^{\circ}$), the KAT-7 RC extends to $\sim$ 1 kpc further out than the ATCA rotation curve. 

3. The observationally motivated DM ISO model reproduces very well the observed RC while the NFW model gives a much poorer fit, especially in the inner parts. This confirms previous results that NGC 6822 has a cored and not a cuspy DM halo. Our derived best fit M/L, 0.12 $\pm$ 0.01, is consistent with the literature value of 0.10 $\pm$ 0.13. The small M/L ratio of NGC 6822 shows that the stellar distribution has no significant contribution to the total mass of the galaxy. The MOND fit fails to reproduce the observed mass distribution in NGC 6822. 

5. The cloud growth criterion produces a better explanation of star formation in NGC 6822 than the Toomre-Q criterion. This shows that local shear rate could be a key player in cloud formation for irregular galaxies such as NGC 6822.

\section*{ACKNOWLEDGEMENT}
We thank the entire SKA SA team for allowing us to obtain scientific data during the commissioning phase of KAT-7. This research is supported by the South 
African Research Chairs Initiative (SARCHI) of the Department of Science and Technology (DST), the Square Kilometer Array South Africa (SKA SA) and the 
National Research Foundation (NRF)
%%%%%%%%%%%%%%%%%%%%%%%%%%%%%%%%%%%%%%%%%%%%%%%%%%



%\bibliographystyle{mnras}

\bibliographystyle{mn2e}
%\bibliography{main} 
%%%%%%%%%%%%%%%%%%%%%%%%%%%%%%%%%%%%%%%%%%%%%%%%%%

%%%%%%%%%%%%%%%%% APPENDICES %%%%%%%%%%%%%%%%%%%%%

%\appendix



%%%%%%%%%%%%%%%%%%%%%%%%%%%%%%%%%%%%%%%%%%%%%%%%%%

% Don't change these lines
\bsp	% typesetting comment
\label{lastpage}
\end{document}